\documentclass[%
 reprint,
superscriptaddress,
 amsmath,amssymb,
 aps,
 prl,
]{revtex4-1}
\usepackage{natbib}
\usepackage{graphicx}
\usepackage{float}
\usepackage{graphicx}
\usepackage{dcolumn}
\usepackage{bm}
\usepackage{verbatim}
\usepackage[dvipsnames]{xcolor}
\usepackage{hyperref}
\usepackage[mathlines]{lineno}
\usepackage{amsmath}
\begin{document}

\preprint{APS/123-QED}

\title{High-quality multi-wavelength quantum light sources on silicon nitride micro-ring chip}
\author{Yun-Ru Fan}
\affiliation{Institute of Fundamental and Frontier Sciences, University of Electronic Science and Technology of China, Chengdu 610054, China}
\author{Chen Lyu}
\affiliation{Institute of Fundamental and Frontier Sciences, University of Electronic Science and Technology of China, Chengdu 610054, China}
\author{Chen-Zhi Yuan}
\affiliation{Institute of Fundamental and Frontier Sciences, University of Electronic Science and Technology of China, Chengdu 610054, China}
\author{Guang-Wei Deng}
\affiliation{Institute of Fundamental and Frontier Sciences, University of Electronic Science and Technology of China, Chengdu 610054, China}
\author{Zhi-Yuan Zhou}
\affiliation{CAS Key Laboratory of Quantum Information, University of Science and Technology of China, Hefei 230026, China}
\author{Yong Geng}
\affiliation{Key Lab of Optical Fiber Sensing and Communication Networks, University of Electronic Science and Technology of China, Chengdu 611731, China}
\author{Hai-Zhi Song}
\affiliation{Institute of Fundamental and Frontier Sciences, University of Electronic Science and Technology of China, Chengdu 610054, China}
\affiliation{Southwest Institute of Technical Physics, Chengdu 610041, China}
\author{You Wang}
\affiliation{Institute of Fundamental and Frontier Sciences, University of Electronic Science and Technology of China, Chengdu 610054, China}
\affiliation{Southwest Institute of Technical Physics, Chengdu 610041, China}
\author{Yan-Feng Zhang}
\affiliation{School of Electronics and Information Technology, Sun Yat-sen University, Guangzhou 510275, China}
\author{Rui-Bo Jin}
\affiliation{Hubei Key Laboratory of Optical Information and Pattern Recognition, Wuhan Institute of Technology, Wuhan 430205, China}
\author{Heng Zhou}
\affiliation{Key Lab of Optical Fiber Sensing and Communication Networks, University of Electronic Science and Technology of China, Chengdu 611731, China}
\author{Li-Xing You}
\affiliation{Shanghai Institute of Microsystem and Information Technology, Chinese Academy of Sciences, Shanghai 200050, China}
\author{Guang-Can Guo}
\affiliation{Institute of Fundamental and Frontier Sciences, University of Electronic Science and Technology of China, Chengdu 610054, China}
\affiliation{CAS Key Laboratory of Quantum Information, University of Science and Technology of China, Hefei 230026, China}
\author{Qiang Zhou}
\email{zhouqiang@uestc.edu.cn}
\affiliation{Institute of Fundamental and Frontier Sciences, University of Electronic Science and Technology of China, Chengdu 610054, China}
\affiliation{CAS Key Laboratory of Quantum Information, University of Science and Technology of China, Hefei 230026, China}
\date{\today}
\begin{abstract}
Multi-wavelength quantum light sources, especially at telecom band, are extremely desired in quantum information technology.~Despite recent impressive advances, such a quantum light source with high quality remains challenging.~Here we demonstrate a multi-wavelength quantum light source using a silicon nitride micro-ring with a free spectral range of 200 GHz.~The generation of eight pairs of correlated photons is ensured in a wavelength range of 25.6 nm.~With device optimization and noise-rejecting filters, our source enables the generation of heralded single-photons - at a rate of 62 kHz with $g^{(2)}_{h}(0)=0.014\pm0.001$, and the generation of energy-time entangled photons - with a visibility of $99.39\pm 0.45\%$ in the Franson interferometer.~These results, at room temperature and telecom wavelength, in a CMOS compatible platform, represent an important step towards integrated quantum light devices for the quantum networks.
\end{abstract}
\maketitle

\textit{Introduction.}—Quantum light sources are elementary building blocks for realizing quantum networks\cite{bouwmeester1997, lloyd1996universal, mitchell2004, chen2021, liu2019solid, chen2021bright, lu2021quantum, wei2022towards}, which are essential for both fundamental insights and down-to-earth applications.~Multi-wavelength quantum light sources have proven to be an indispensable component to establish a spectrally multiplexed quantum repeater\cite{sinclair2014spectral}.~Furthermore, they are also beneficial to photonic quantum technologies\cite{wang2018multidimensional, cerf2002security, bouchard2017high, huang2022experimental,paesani2021scheme, peters2021machine, li2020metalens, erhard2020advances}, such as improving the performance of quantum communication, enabling photonic quantum computation, and enhancing the sensitivity of quantum metrology.~Recently, integrated quantum light sources based on cavity-enhanced devices have leveraged strengths of multi-wavelength properties benefited from high Q-factor, small mode volume, and flexible dispersion engineering\cite{caspani2016, chang2021}, which offers a versatile platform with numerous superiorities, namely compatibility, scalability, and reconfigurability.~Based on processes of spontaneous parametric down-conversion (SPDC) and spontaneous four-wave mixing (SFWM),~integrated quantum light sources have been developed in the second-order\cite{ma2020ultrabright, luo2017chip, xu2021spectrally, guo2017parametric} and third-order nonlinear optical devices\cite{ongsilicon2011, kumarsilicon2013, kumarsilicon2015, jiang2015silicon, ramelow2015, jaramillo2017, imany2018, lu2019, samara2019, li2019tunable, lu2022bayesian, yin2021, samara2021entanglement,reimer2014, reimer2016science, reimer2019,liu2020photolithography}.~For instance, five pairs of correlated photons with a periodically poled lithium niobate micro-ring resonator (MRR)\cite{ma2020ultrabright} and multiphoton entangled quantum states with Hydex glass MRR\cite{reimer2016science} have been realized, which have great potential in applications.~Despite these impressive results, multi-wavelength quantum light sources at telecom band with high quality, such as generation rates and entanglement properties, are still with challenges, thus hinder their full potential.~To that end, the quantum correlation property, the noise property and photon input-output mechanism, and the entanglement manipulation of such sources should be further investigated.

Here we demonstrate a multi-wavelength quantum light source at telecom band on silicon nitride MRR chip.~The generation of eight pairs of correlated photons with high quality is obtained,~namely high photon generation rate, coincidence-to-accidental ratio (CAR) and single-photon purity, in a range of 25.6 nm with a free spectral range of 200 GHz.~The generation of heralded single photons and energy-time entanglement for each wavelength-pair are achieved with $g^{(2)}_{h}$(0)=0.014$\pm$0.001 at a rate of 62 kHz and a Franson interference visibility of 99.39$\pm$0.45\%, respectively. Our demonstrations suggest that the integrated quantum light source based on MRR opens promising perspectives in the future quantum networks.

\begin{figure*}
    \centering
    \includegraphics[width=17 cm]{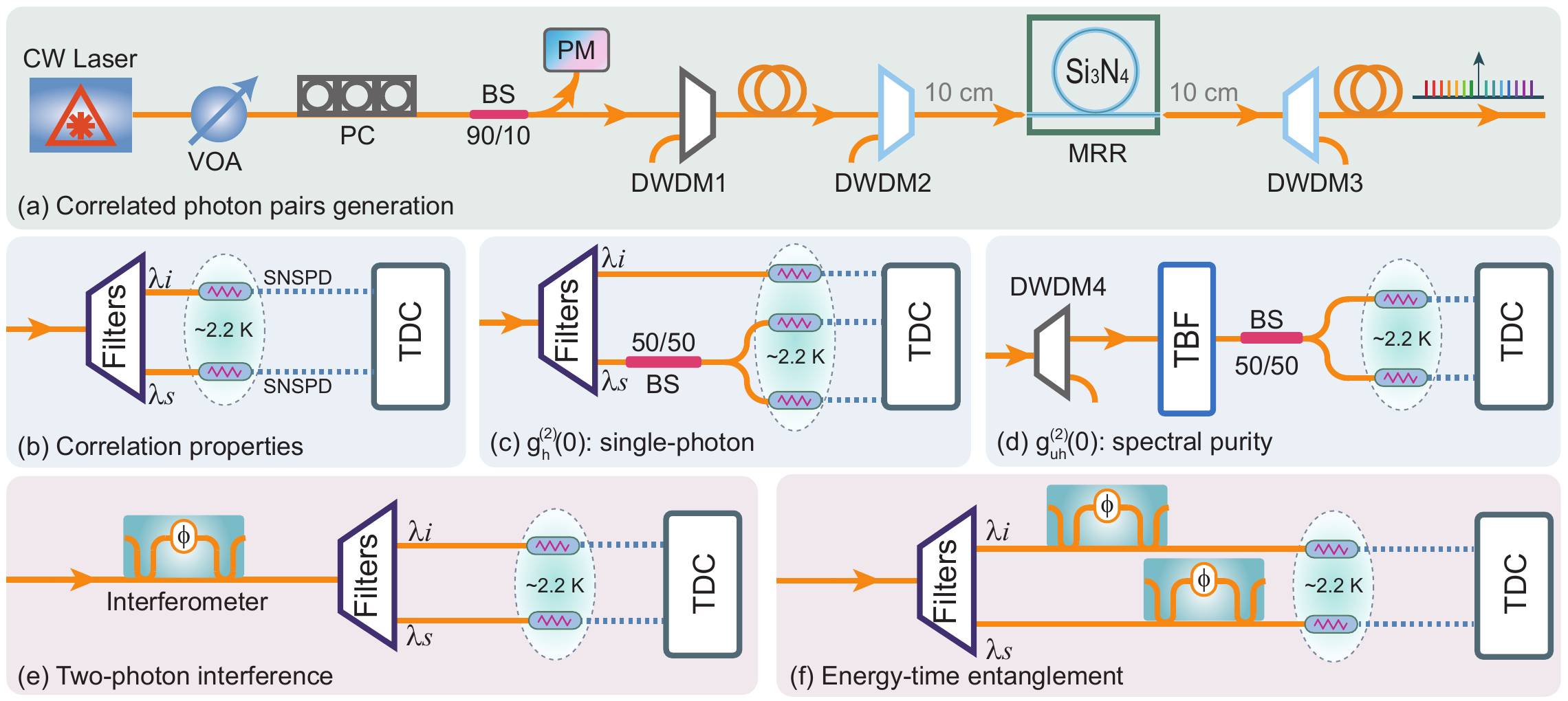}
    \caption{Experimental setups for the generation and characterization of our quantum light sources. (a) Correlated photon pairs generation, (b) Correlation properties, (c) Heralded single-photon source, (d) Spectral purity, (e) Two-photon interference, and (f) Energy-time entanglement. VOA: variable optical attenuator; PC: polarization controller; BS: beam splitter; PM: power meter; DWDM: dense-wavelength division multiplexer; MRR: micro-ring resonator; SNSPD: superconducting nanowire single photon detector; TDC: time-to-digital convertor; TBF: tunable bandpass filter.
    }
    \label{fig:Fig1}
\end{figure*}

\textit{Device design and characterization.}—We design and fabricate four groups of silicon nitride MRRs with different coupling conditions.~With our analysis of photon input-output mechanism and device selection, an optimized over-coupling MRR with the Q-factor of 1.0$\times$10$^6$ and the free spectral range of $\sim$200 GHz is employed for developing the quantum light source.~See details in Supplementary Materials Note1 and Note2.

\textit{Multi-wavelength correlated photon pairs.}—The experimental setup for the generation of multi-wavelength correlated/entangled photon pairs is shown in Fig.~{\ref{fig:Fig1}}(a).~An important feature of our source is that the MRR is connected with noise-rejecting filters with 10 cm fiber-pigtails,~thus mitigating the noise photons from other fiber-pigtails.~The noise analyses are shown in Supplementary Materials Note3.~As illustrated in Fig.~{\ref{fig:Fig1}}(b),~quantum correlations are measured between a typical photon pairs at 1550.1 nm (C34) and 1531.0 nm (C58),~with a pump wavelength of 1540.5 nm (C46).
\begin{figure*}
    \centering
    \includegraphics[width=17 cm]{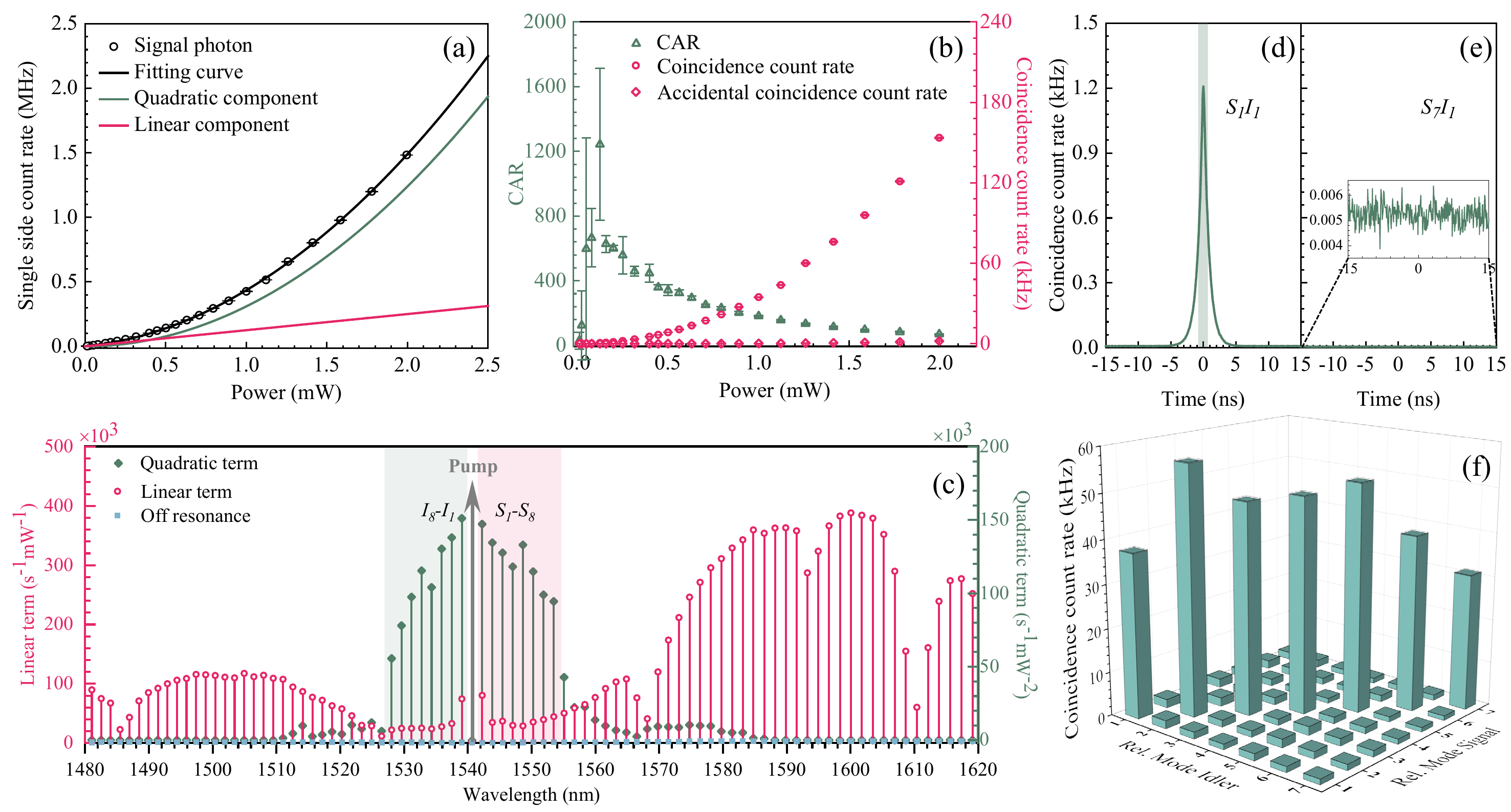}
    \caption{Multi-wavelength correlated photon pairs.~(a) Single side photon count rate of signal photons at 1550.1 nm versus pump. (b) Coincidence count rate, accidental coincidence count rate, and the calculated CAR versus pump. (c) Spectrum of the correlated photons and noise photons. (d) Correlations for \textit{$S_1I_1$}. (e) Correlations for \textit{$S_7I_1$}. (f) Joint spectral intensity.}
    \label{fig:Fig2}
\end{figure*}

Figure~{\ref{fig:Fig2}}(a) shows the measured single side count rate on the resonance of 1550.1 nm under different pump power levels.~Circles are the experimental results, and the black line is the $aP{_p}^2+bP_p$ fitting, in which the part of $aP{_p}^2$ is the contribution of correlated photons while $bP_p$ is the contribution of noise photons\cite{samara2019}. The quantum correlation property is further characterized by measuring the coincidence and accidental coincidence count rate and the ratios between them (CAR). In these measurements, the coincidence window is 1.6 ns which corresponds to the full width at half maximum of the coincidence histogram as shown in Fig.~{\ref{fig:Fig2}}(d).~The measured results are given in Fig.~{\ref{fig:Fig2}}(b), in which a CAR of $1243\pm469$ is achieved with a signal photon count rate of 24 kHz.

Single side photon count rates at different resonance wavelengths are measured by using a tunable bandpass filter (TBF, XTA-50, EXFO).~Contributions of correlated photons and noise photons are obtained by the $aP{_p^2}+bP_p$ fitting of the measured results, as summarized in Fig.~{\ref{fig:Fig2}}(c).~Although photons are generated on all the measured resonances from 1480 nm to 1620 nm (see details in Supplementary Materials Note3),~only eight pairs of resonances,~i.e.~sixteen resonances in a range of 25.6 nm,~have contributions from correlated photons (green lines,~labeled as \textit{$S_nI_n$},~where $n=1, 2, ...8$),~while other resonances are dominated by noise photons.~The blue lines show the Raman noise in MRR with off-resonance pump case, which are negligible thanks to the noise-rejecting design.~The multi-wavelength property of the obtained eight pairs of correlated photons is also observed by measuring the coincidence among these wavelengths.~Figure~{\ref{fig:Fig2}}(d) shows a typical coincidence histogram with wavelength correlated,~i.e.~\textit{$S_1I_1$},~while Fig.~{\ref{fig:Fig2}}(e) is the coincidence measurement with wavelength uncorrelated,~i.e.~\textit{$S_7I_1$}.~The coincidences among seven pairs are measured and shown in Fig.~{\ref{fig:Fig2}}(f), which shows that coincidence is measured only with wavelength correlated ones - the $S_8I_8$ one is outside our measurements.
\begin{figure}
    \centering
    \includegraphics[width=8.5 cm]{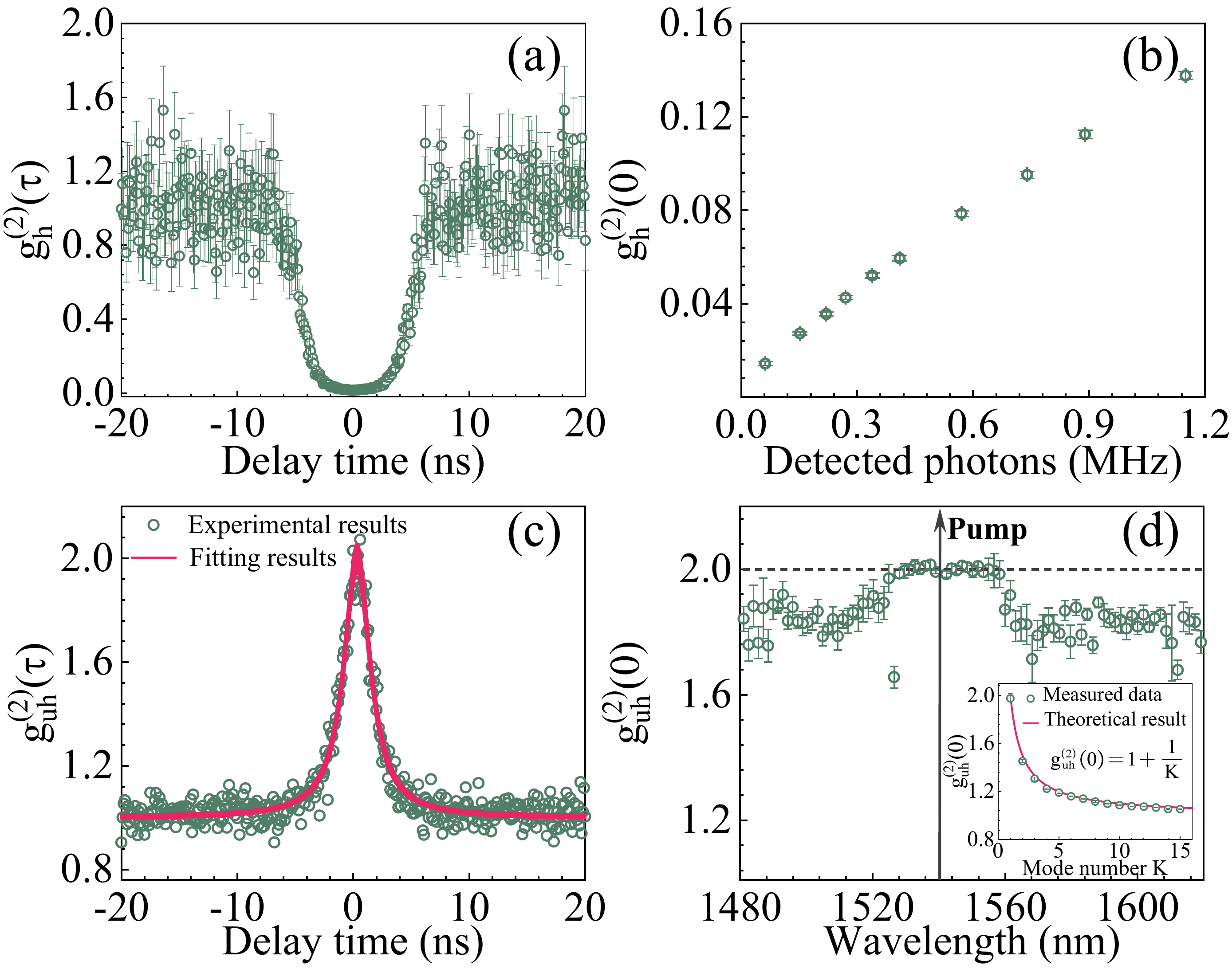}
    \caption{Multi-wavelength heralded single-photon source.~(a) Measured heralded second-order auto-correlation $g^{(2)}_{h}(\tau)$ with $g^{(2)}_{h}(0)=0.014\pm0.001$.~(b) $g^{(2)}_{h}(0)$ as the function of the detected idler photons.~(c) Measured signal-signal (idler-idler) second-order auto-correlation $g^{(2)}_{uh}(\tau)$, with $g^{(2)}_{uh}(0)=1.997\pm0.017$).~(d) $g^{(2)}_{uh}(0)$ at different resonant wavelengths. The inset gives the result of $g^{(2)}_{uh}(0)$ with effective mode number from 1 to 15.}
    \label{fig:Fig3}
\end{figure}

\textit{Heralded single-photon source.}—To characterize the performance of our device as a multi-wavelength single-photon source,~we measure the heralded second-order auto-correlation function $g^{(2)}_{h}(\tau)$ to evaluate the single-photon purity,~as well as the signal-signal (idler-idler) second-order auto-correlation function $g^{(2)}_{uh}(\tau)$ to estimate the spectral purity by using the Hanbury Brown and Twiss (HBT) setup\cite{brown1956, steudle2012}.~As shown in Fig.~{\ref{fig:Fig1}}(c),~we use the detection of idler photons to herald the presence of correlated signal photons (heralded photons),~which are split by a 50/50 beam splitter (BS) and labeled as signal1 and signal2,~respectively.~The heralded second-order auto-correlation function $g^{(2)}_{h}(\tau)$ can be calculated by\cite{ma2020ultrabright},
\begin{equation}
\label{eq:1}
g_{h}^{\left( 2\right)}\left(\tau\right)=\frac{N_{123}\left(\tau \right)\times N_1}{N_{12}\times N_{13}\left(\tau\right)},
\end{equation}
where $N_1$, $N_{12}$ ($N_{13}(\tau)$), $N_{123}(\tau)$, and $\tau$ are the idler, two-fold coincidence, three-fold coincidence count rate, and the relative delay between signal1 and signal2, respectively.

Figure~{\ref{fig:Fig3}}(a) shows a measured curve of $g^{(2)}_{h}(\tau)$ with \textit{$S_6I_6$}.~A distinct anti-bunching dip is observed with $g^{(2)}_{h}(0)= 0.014\pm0.001$ at an idler photon count rate of 62 kHz,~which is smaller than the limit for classical correlation and confirms the quantum nature of the heralded single-photon source.~The $g^{(2)}_{h}(0)$ values with different idler photon count rates (heralding photons) are shown in Fig.~{\ref{fig:Fig3}}(b).~The detected idler photon count rate reaches 1.15 MHz with a $g^{(2)}_{h}(0)$ of $0.138\pm0.002$.~The $g^{(2)}_{h}(0)$ values of another six sources are given in Table~{\ref{tab:tableI}}.
\begin{table*}
\caption{\label{tab:tableI} Experimental results of heralded single-photon sources and energy-time entanglement.}
\begin{ruledtabular}
\begin{tabular}{cccccc}
Channels\footnote{ITU Wavelength} & Detected heralding count rate & $g^{2}_{h}(\tau=0)$&$V^{(f)}$&$V^{(uf)}$ & Violation of Bell's inequality\\\hline
C32$\&$C60 &245 kHz & 0.049$\pm$0.002&96.28$\pm$0.42\%& 97.33$\pm$0.49\%&54 Std.\\
C34$\&$C58 &339 kHz & 0.052$\pm$0.001&99.39$\pm$0.45\%& 99.55$\pm$0.33\%&87 Std.\\
C36$\&$C56 &300 kHz & 0.051$\pm$0.001&98.31$\pm$0.84\%& 100.00$\pm$0.02\%&-\\
C38$\&$C54 &290 kHz & 0.045$\pm$0.001&99.97$\pm$0.14\%& 99.97$\pm$0.10\%&292 Std.\\
C40$\&$C52 &315 kHz & 0.058$\pm$0.001&99.61$\pm$0.58\%& 100.00$\pm$0.01\%&73 Std.\\
C42$\&$C50 &400 kHz & 0.064$\pm$0.002&99.30$\pm$0.77\%& 99.99$\pm$0.04\%&-\\
C44$\&$C48 &353 kHz & 0.079$\pm$0.001&99.08$\pm$0.86\%& 98.91$\pm$0.52\%&54 Std.\\
\end{tabular}
\end{ruledtabular}
\end{table*}

The spectral purity properties of generated photons are given in Fig.~{\ref{fig:Fig1}}(d). The photons are selected with DWDM4 and the TBF, then sent into a 50/50 BS.~Figure~{\ref{fig:Fig3}}(c) shows the measured auto-correlation $g^{(2)}_{uh}(\tau)$ for the photons at 1550.1 nm.~A $g^{(2)}_{uh}(0)$ of $1.997\pm0.017$ is obtained, which is close to the ideal value of $g^{(2)}_{uh}(0)$=2 for single-mode thermal photon, indicating a high spectral purity of our source. The dots are the measured data and the lines are the fitting result with 1000-time Monte Carlo method. With the same method, Fig.~{\ref{fig:Fig3}}(d) gives the $g^{(2)}_{uh}(0)$ values for all resonant wavelengths, which are close to 2 for wavelengths with correlated photons while ~1.7 for ones without correlated photons. For the multi-mode thermal photon case, we measure the $g^{(2)}_{uh}(0)$ with a variety of effective mode number from 1 to 15 by adding resonance peaks with a programmable filter, as shown in the inset of Fig.~{\ref{fig:Fig1}}(d). The dots are the measured results and the line is the theoretical result of $g^{(2)}_{uh}(0)=1+1/K$, where $K$ is the effective mode number.

\textit{Two-photon interference and energy-time entanglement.}—The generated photon pairs in our device are expected to be a multi-wavelength entangled quantum light source, which can be measured and manipulated by two-photon interference\cite{franson1989bell, kwiat1993high}. Figure~{\ref{fig:Fig1}}(e) shows measured two-photon interference using a folded Franson interferometer, in which both signal and idler photons propagate along the same unbalanced fiber Michelson interferometer (UMI) with a delay time of 10 ns.~It is worth noting that an attenuated CW laser is also injected into the UMI for phase stabilization based on feedback control.~As shown in Fig.~{\ref{fig:Fig4}}(a), red and green dots and lines are the measured two-photon and single-photon interference and the fitting results, respectively.~A raw visibility of $97.3\pm0.6\%$ is obtained for the two-photon interference, clearly in excess of the threshold value of $70.7\%$ required to violate adapted Clauser-Horne-Shimony-Holt (CHSH) Bell inequalities. With background subtraction, the visibility~$V^{(f)}$ reaches $99.39\pm0.45\%$.
\begin{figure}
    \centering
    \includegraphics[width=8.5 cm]{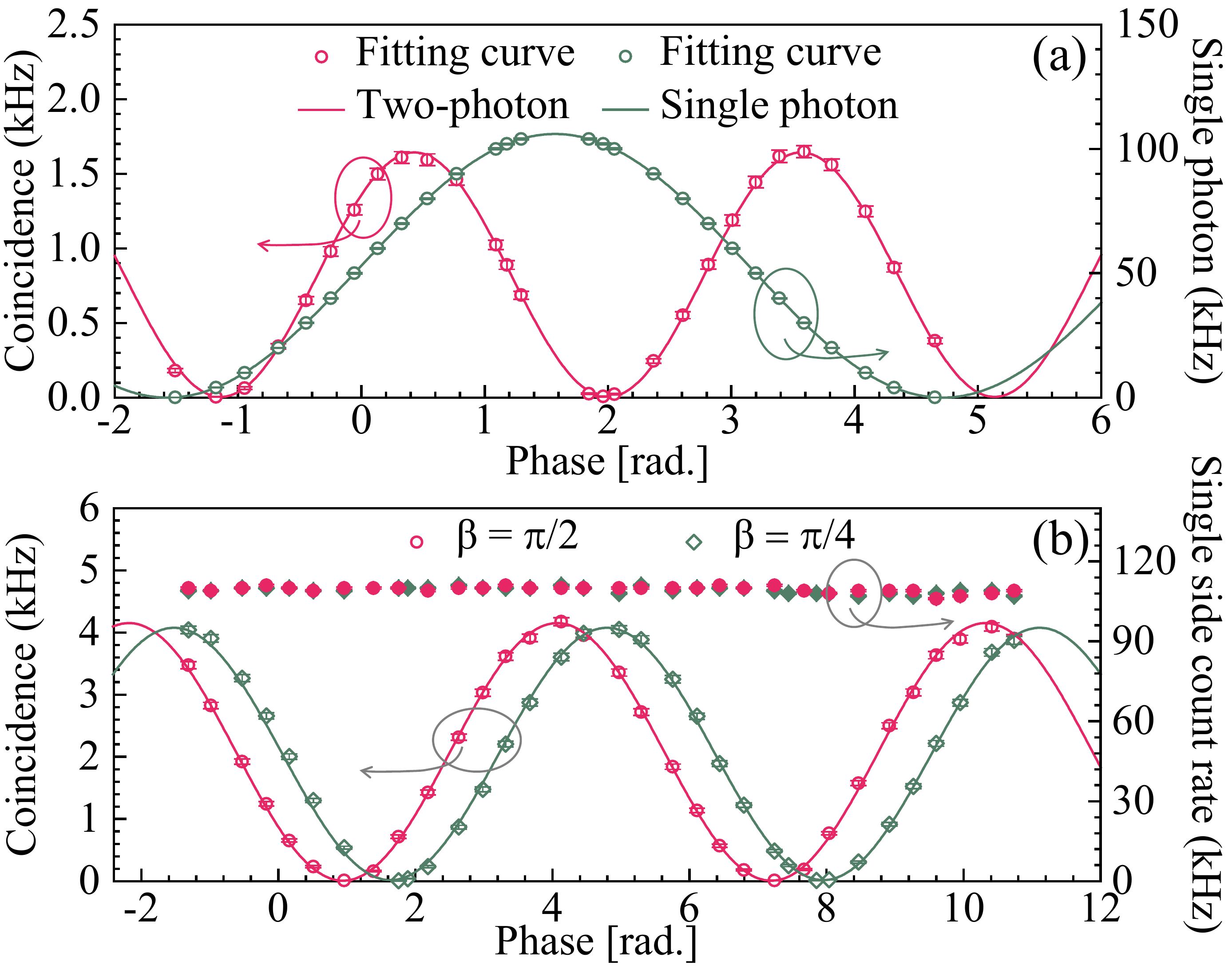}
    \caption{Characterization of energy-time entanglement. (a) Results of two-photon interference (red dots) and single-photon interference (green dots) with folded Franson interferometer. (b) Results of Franson interference for $\beta$=$\pi/2$ (red dots) and $\beta$=$\pi/4$ (green dots) with unfolded Franson interferometer. The right vertical axis presents the single side count rate at signal channel.}
    \label{fig:Fig4}
\end{figure}

Figure~{\ref{fig:Fig1}}(f) shows the setup for measuring the energy-time entanglement with the unfolded Franson interferometer.~The signal and idler photons are separated by DWDM filters, and then injected into two identical UMIs with an additional phase difference of $\alpha$ or $\beta$.~As shown in Fig.~{\ref{fig:Fig4}}(b), energy-time entanglement properties are confirmed by the sinusoidal variation of coincidences with phase.~The visibility is $V^{(uf)}=99.55\pm0.33\%$ (red circles) and $98.66\pm0.39\%$ (green circles) with background subtraction for two phases of $\beta=\pi/2$ and $\beta=\pi/4$, respectively.~The calculated S parameter is S=2.816$\pm$0.010 for $\beta=\pi/2$ and S=2.791$\pm$0.011 for $\beta=\pi/4$, which shows a violation of the CHSH Bell inequality by more than 87 and 71 \cite{marcikic2004}, respectively.~In both cases, the single side count rates keep constant. The energy-time entanglement properties of another six entangled photon pairs are listed in Table~{\ref{tab:tableI}}.

\textit{Discussions and summary.}—An important feature of our demonstration is the multi-wavelength property of quantum light sources with high quality. With the same silicon nitride MRR device, its classical counterpart, i.e. the Kerr-frequency comb source, can generate frequency comb in a bandwidth of $\sim$100 nm\cite{del2007optical, kippenberg2011microresonator, herr2012universal,zhou2019soliton, geng2022coherent}.~The bandwidth of our quantum light source is much smaller than that of the classical one, which could be attributed to their generation processes.~For Kerr-frequency comb, the high pump power level in cavity enables cascaded Kerr and Raman scattering nonlinear processes,~resulting in the generation of the optical frequency comb in a large wavelength range.~While the quantum light source is only generated from the SFWM process at low pump power level, which has a much smaller wavelength range due to the sophisticated phase-matching requirement.~We would suggest a further avenue to improve the number of frequency combs and the generation rates.~The number of frequency combs could be increased through dispersion engineering of the device.~It could also be achieved, for example, by decreasing the FSR of the resonator.~To advance a higher generation rate with the same pump power, the key is to reach a higher $Q_i$ by reducing the scattering and bending losses in the fabrication.~Given the trade-off between the generation in and emission from the ring cavity, the $Q_e$ should be selected for an over-coupling regime by designing the gap width between the bus waveguide and the ring cavity.~Alternatively, this trade-off can also be overcome by utilizing dual asymmetric Mach-Zehnder interferometers\cite{wu2020bright} or employing a $\mathcal{PT}$-symmetric coupled dual-ring structure\cite{peng2014parity, lu2022parity},~in which the life times of pump, signal, and idler photons can be manipulated individually.~The overall rate can be further improved by using filters with low losses which can be integrated on chip\cite{perez2017optical,yu2022spectrally}.

The multi-wavelength nature of cavity-enhanced resonators can pave the way to explore spectrally multiplexed single-photon sources.~Utilizing advanced technologies like single-photon frequency shifting\cite{yu2022spectrally} and broadband single-photon spectrometer~-~frequency-resolved single-photon detection\cite{cheng2019broadband}, heralded single-photon sources with higher efficiency and deterministic manner can be achieved which addresses the limitation of heralded single-photon rate and single-photon purity.~Towards the future large-scale quantum internet, quantum information encoded in spectral domain represents a potential approach in quantum communication, quantum simulation and computation, and quantum metrology. For instance, the quantum walk in frequency domain could be implemented with comb-like states generated in cavities, and thus sufficiently accomplishing complex quantum algorithms with spectral manipulation\cite{imany2020probing}. Our results suggest that micro-cavity-based photon sources and their natural combs open new avenues with capability of multi-channel parallel quantum information processing in frequency-domain with integrated photonics. Furthermore, our demonstrations provide an important source for multi-user quantum key distribution protocol\cite{liu202240, wen2022realizing} and for fully connected entanglement-based quantum networks\cite{wengerowsky2018entanglement, joshi2020trusted}.

In summary, we have demonstrated high-quality multi-wavelength quantum light sources on a silicon nitride MRR chip.~Enabled by the natural resonance characteristics,~the generation of eight-wavelength photon pairs has been achieved simultaneously in a range of 25.6 nm with an FSR of 200 GHz.~The quantum signature has been confirmed by the generation of heralded single-photon with a second-order auto-correlation value of $g^{(2)}_{h}(0)=0.014\pm0.001$, and by the energy-time entanglement with a typical visibility of $99.39\pm0.45\%$.~The large-scale integration potential and stability of silicon nitride with CMOS compatible fabrication would also motivate the use of multi-wavelength quantum light sources in practical applications.~These results suggest that our high-quality multi-wavelength quantum light sources represent a promising avenue for the developing of quantum information technologies.

\begin{acknowledgments}
This work was supported by the National Key Research and Development Program of China (Nos.~2018YFA0307400, 2018YFA0306102), National Natural Science Foundation of China (Nos.~61775025, 91836102, U19A2076, 62005039),~Sichuan Science and Technology Program (Nos.~2021YFSY0063, 2021YFSY0062, 2021YFSY0064, 2021YFSY0065, 2021YFSY0066),~Innovation Program for Quantum Science and Technology (No.~2021ZD0301702).~The authors acknowledge LiGenTec SA for device fabrication.
\end{acknowledgments}
\bibliography{referencemain}

\begin{thebibliography}{65}%
\makeatletter
\providecommand \@ifxundefined [1]{%
 \@ifx{#1\undefined}
}%
\providecommand \@ifnum [1]{%
 \ifnum #1\expandafter \@firstoftwo
 \else \expandafter \@secondoftwo
 \fi
}%
\providecommand \@ifx [1]{%
 \ifx #1\expandafter \@firstoftwo
 \else \expandafter \@secondoftwo
 \fi
}%
\providecommand \natexlab [1]{#1}%
\providecommand \enquote  [1]{``#1''}%
\providecommand \bibnamefont  [1]{#1}%
\providecommand \bibfnamefont [1]{#1}%
\providecommand \citenamefont [1]{#1}%
\providecommand \href@noop [0]{\@secondoftwo}%
\providecommand \href [0]{\begingroup \@sanitize@url \@href}%
\providecommand \@href[1]{\@@startlink{#1}\@@href}%
\providecommand \@@href[1]{\endgroup#1\@@endlink}%
\providecommand \@sanitize@url [0]{\catcode `\\12\catcode `\$12\catcode
  `\&12\catcode `\#12\catcode `\^12\catcode `\_12\catcode `\%12\relax}%
\providecommand \@@startlink[1]{}%
\providecommand \@@endlink[0]{}%
\providecommand \url  [0]{\begingroup\@sanitize@url \@url }%
\providecommand \@url [1]{\endgroup\@href {#1}{\urlprefix }}%
\providecommand \urlprefix  [0]{URL }%
\providecommand \Eprint [0]{\href }%
\providecommand \doibase [0]{http://dx.doi.org/}%
\providecommand \selectlanguage [0]{\@gobble}%
\providecommand \bibinfo  [0]{\@secondoftwo}%
\providecommand \bibfield  [0]{\@secondoftwo}%
\providecommand \translation [1]{[#1]}%
\providecommand \BibitemOpen [0]{}%
\providecommand \bibitemStop [0]{}%
\providecommand \bibitemNoStop [0]{.\EOS\space}%
\providecommand \EOS [0]{\spacefactor3000\relax}%
\providecommand \BibitemShut  [1]{\csname bibitem#1\endcsname}%
\let\auto@bib@innerbib\@empty
\bibitem [{\citenamefont {Bouwmeester}\ \emph {et~al.}(1997)\citenamefont
  {Bouwmeester}, \citenamefont {Pan}, \citenamefont {Mattle}, \citenamefont
  {Eibl}, \citenamefont {Weinfurter},\ and\ \citenamefont
  {Zeilinger}}]{bouwmeester1997}%
  \BibitemOpen
  \bibfield  {author} {\bibinfo {author} {\bibfnamefont {D.}~\bibnamefont
  {Bouwmeester}}, \bibinfo {author} {\bibfnamefont {J.-W.}\ \bibnamefont
  {Pan}}, \bibinfo {author} {\bibfnamefont {K.}~\bibnamefont {Mattle}},
  \bibinfo {author} {\bibfnamefont {M.}~\bibnamefont {Eibl}}, \bibinfo {author}
  {\bibfnamefont {H.}~\bibnamefont {Weinfurter}}, \ and\ \bibinfo {author}
  {\bibfnamefont {A.}~\bibnamefont {Zeilinger}},\ }\href@noop {} {\bibfield
  {journal} {\bibinfo  {journal} {Nature}\ }\textbf {\bibinfo {volume} {390}},\
  \bibinfo {pages} {575} (\bibinfo {year} {1997})}\BibitemShut {NoStop}%
\bibitem [{\citenamefont {Lloyd}(1996)}]{lloyd1996universal}%
  \BibitemOpen
  \bibfield  {author} {\bibinfo {author} {\bibfnamefont {S.}~\bibnamefont
  {Lloyd}},\ }\href@noop {} {\bibfield  {journal} {\bibinfo  {journal}
  {Science}\ }\textbf {\bibinfo {volume} {273}},\ \bibinfo {pages} {1073}
  (\bibinfo {year} {1996})}\BibitemShut {NoStop}%
\bibitem [{\citenamefont {Mitchell}\ \emph {et~al.}(2004)\citenamefont
  {Mitchell}, \citenamefont {Lundeen},\ and\ \citenamefont
  {Steinberg}}]{mitchell2004}%
  \BibitemOpen
  \bibfield  {author} {\bibinfo {author} {\bibfnamefont {M.~W.}\ \bibnamefont
  {Mitchell}}, \bibinfo {author} {\bibfnamefont {J.~S.}\ \bibnamefont
  {Lundeen}}, \ and\ \bibinfo {author} {\bibfnamefont {A.~M.}\ \bibnamefont
  {Steinberg}},\ }\href@noop {} {\bibfield  {journal} {\bibinfo  {journal}
  {Nature}\ }\textbf {\bibinfo {volume} {429}},\ \bibinfo {pages} {161}
  (\bibinfo {year} {2004})}\BibitemShut {NoStop}%
\bibitem [{\citenamefont {Chen}\ \emph
  {et~al.}(2021{\natexlab{a}})\citenamefont {Chen}, \citenamefont {Zhang},
  \citenamefont {Chen}, \citenamefont {Cai}, \citenamefont {Liao},
  \citenamefont {Zhang}, \citenamefont {Chen}, \citenamefont {Yin},
  \citenamefont {Ren}, \citenamefont {Chen} \emph {et~al.}}]{chen2021}%
  \BibitemOpen
  \bibfield  {author} {\bibinfo {author} {\bibfnamefont {Y.-A.}\ \bibnamefont
  {Chen}}, \bibinfo {author} {\bibfnamefont {Q.}~\bibnamefont {Zhang}},
  \bibinfo {author} {\bibfnamefont {T.-Y.}\ \bibnamefont {Chen}}, \bibinfo
  {author} {\bibfnamefont {W.-Q.}\ \bibnamefont {Cai}}, \bibinfo {author}
  {\bibfnamefont {S.-K.}\ \bibnamefont {Liao}}, \bibinfo {author}
  {\bibfnamefont {J.}~\bibnamefont {Zhang}}, \bibinfo {author} {\bibfnamefont
  {K.}~\bibnamefont {Chen}}, \bibinfo {author} {\bibfnamefont {J.}~\bibnamefont
  {Yin}}, \bibinfo {author} {\bibfnamefont {J.-G.}\ \bibnamefont {Ren}},
  \bibinfo {author} {\bibfnamefont {Z.}~\bibnamefont {Chen}},  \emph {et~al.},\
  }\href@noop {} {\bibfield  {journal} {\bibinfo  {journal} {Nature}\ }\textbf
  {\bibinfo {volume} {589}},\ \bibinfo {pages} {214} (\bibinfo {year}
  {2021}{\natexlab{a}})}\BibitemShut {NoStop}%
\bibitem [{\citenamefont {Liu}\ \emph {et~al.}(2019)\citenamefont {Liu},
  \citenamefont {Su}, \citenamefont {Wei}, \citenamefont {Yao}, \citenamefont
  {Silva}, \citenamefont {Yu}, \citenamefont {Iles-Smith}, \citenamefont
  {Srinivasan}, \citenamefont {Rastelli}, \citenamefont {Li} \emph
  {et~al.}}]{liu2019solid}%
  \BibitemOpen
  \bibfield  {author} {\bibinfo {author} {\bibfnamefont {J.}~\bibnamefont
  {Liu}}, \bibinfo {author} {\bibfnamefont {R.}~\bibnamefont {Su}}, \bibinfo
  {author} {\bibfnamefont {Y.}~\bibnamefont {Wei}}, \bibinfo {author}
  {\bibfnamefont {B.}~\bibnamefont {Yao}}, \bibinfo {author} {\bibfnamefont
  {S.~F. C.~d.}\ \bibnamefont {Silva}}, \bibinfo {author} {\bibfnamefont
  {Y.}~\bibnamefont {Yu}}, \bibinfo {author} {\bibfnamefont {J.}~\bibnamefont
  {Iles-Smith}}, \bibinfo {author} {\bibfnamefont {K.}~\bibnamefont
  {Srinivasan}}, \bibinfo {author} {\bibfnamefont {A.}~\bibnamefont
  {Rastelli}}, \bibinfo {author} {\bibfnamefont {J.}~\bibnamefont {Li}},  \emph
  {et~al.},\ }\href@noop {} {\bibfield  {journal} {\bibinfo  {journal} {Nature
  nanotechnology}\ }\textbf {\bibinfo {volume} {14}},\ \bibinfo {pages} {586}
  (\bibinfo {year} {2019})}\BibitemShut {NoStop}%
\bibitem [{\citenamefont {Chen}\ \emph
  {et~al.}(2021{\natexlab{b}})\citenamefont {Chen}, \citenamefont {Wei},
  \citenamefont {Zhao}, \citenamefont {Liu}, \citenamefont {Su}, \citenamefont
  {Yao}, \citenamefont {Yu}, \citenamefont {Liu},\ and\ \citenamefont
  {Wang}}]{chen2021bright}%
  \BibitemOpen
  \bibfield  {author} {\bibinfo {author} {\bibfnamefont {B.}~\bibnamefont
  {Chen}}, \bibinfo {author} {\bibfnamefont {Y.}~\bibnamefont {Wei}}, \bibinfo
  {author} {\bibfnamefont {T.}~\bibnamefont {Zhao}}, \bibinfo {author}
  {\bibfnamefont {S.}~\bibnamefont {Liu}}, \bibinfo {author} {\bibfnamefont
  {R.}~\bibnamefont {Su}}, \bibinfo {author} {\bibfnamefont {B.}~\bibnamefont
  {Yao}}, \bibinfo {author} {\bibfnamefont {Y.}~\bibnamefont {Yu}}, \bibinfo
  {author} {\bibfnamefont {J.}~\bibnamefont {Liu}}, \ and\ \bibinfo {author}
  {\bibfnamefont {X.}~\bibnamefont {Wang}},\ }\href@noop {} {\bibfield
  {journal} {\bibinfo  {journal} {Nature Nanotechnology}\ }\textbf {\bibinfo
  {volume} {16}},\ \bibinfo {pages} {302} (\bibinfo {year}
  {2021}{\natexlab{b}})}\BibitemShut {NoStop}%
\bibitem [{\citenamefont {Lu}\ and\ \citenamefont {Pan}(2021)}]{lu2021quantum}%
  \BibitemOpen
  \bibfield  {author} {\bibinfo {author} {\bibfnamefont {C.-Y.}\ \bibnamefont
  {Lu}}\ and\ \bibinfo {author} {\bibfnamefont {J.-W.}\ \bibnamefont {Pan}},\
  }\href@noop {} {\bibfield  {journal} {\bibinfo  {journal} {Nature
  Nanotechnology}\ }\textbf {\bibinfo {volume} {16}},\ \bibinfo {pages} {1294}
  (\bibinfo {year} {2021})}\BibitemShut {NoStop}%
\bibitem [{\citenamefont {Wei}\ \emph {et~al.}(2022)\citenamefont {Wei},
  \citenamefont {Jing}, \citenamefont {Zhang}, \citenamefont {Liao},
  \citenamefont {Yuan}, \citenamefont {Fan}, \citenamefont {Lyu}, \citenamefont
  {Zhou}, \citenamefont {Wang}, \citenamefont {Deng} \emph
  {et~al.}}]{wei2022towards}%
  \BibitemOpen
  \bibfield  {author} {\bibinfo {author} {\bibfnamefont {S.-H.}\ \bibnamefont
  {Wei}}, \bibinfo {author} {\bibfnamefont {B.}~\bibnamefont {Jing}}, \bibinfo
  {author} {\bibfnamefont {X.-Y.}\ \bibnamefont {Zhang}}, \bibinfo {author}
  {\bibfnamefont {J.-Y.}\ \bibnamefont {Liao}}, \bibinfo {author}
  {\bibfnamefont {C.-Z.}\ \bibnamefont {Yuan}}, \bibinfo {author}
  {\bibfnamefont {B.-Y.}\ \bibnamefont {Fan}}, \bibinfo {author} {\bibfnamefont
  {C.}~\bibnamefont {Lyu}}, \bibinfo {author} {\bibfnamefont {D.-L.}\
  \bibnamefont {Zhou}}, \bibinfo {author} {\bibfnamefont {Y.}~\bibnamefont
  {Wang}}, \bibinfo {author} {\bibfnamefont {G.-W.}\ \bibnamefont {Deng}},
  \emph {et~al.},\ }\href@noop {} {\bibfield  {journal} {\bibinfo  {journal}
  {Laser \& Photonics Reviews}\ }\textbf {\bibinfo {volume} {16}},\ \bibinfo
  {pages} {2100219} (\bibinfo {year} {2022})}\BibitemShut {NoStop}%
\bibitem [{\citenamefont {Sinclair}\ \emph {et~al.}(2014)\citenamefont
  {Sinclair}, \citenamefont {Saglamyurek}, \citenamefont {Mallahzadeh},
  \citenamefont {Slater}, \citenamefont {George}, \citenamefont {Ricken},
  \citenamefont {Hedges}, \citenamefont {Oblak}, \citenamefont {Simon},
  \citenamefont {Sohler} \emph {et~al.}}]{sinclair2014spectral}%
  \BibitemOpen
  \bibfield  {author} {\bibinfo {author} {\bibfnamefont {N.}~\bibnamefont
  {Sinclair}}, \bibinfo {author} {\bibfnamefont {E.}~\bibnamefont
  {Saglamyurek}}, \bibinfo {author} {\bibfnamefont {H.}~\bibnamefont
  {Mallahzadeh}}, \bibinfo {author} {\bibfnamefont {J.~A.}\ \bibnamefont
  {Slater}}, \bibinfo {author} {\bibfnamefont {M.}~\bibnamefont {George}},
  \bibinfo {author} {\bibfnamefont {R.}~\bibnamefont {Ricken}}, \bibinfo
  {author} {\bibfnamefont {M.~P.}\ \bibnamefont {Hedges}}, \bibinfo {author}
  {\bibfnamefont {D.}~\bibnamefont {Oblak}}, \bibinfo {author} {\bibfnamefont
  {C.}~\bibnamefont {Simon}}, \bibinfo {author} {\bibfnamefont
  {W.}~\bibnamefont {Sohler}},  \emph {et~al.},\ }\href@noop {} {\bibfield
  {journal} {\bibinfo  {journal} {Physical Review Letters}\ }\textbf {\bibinfo
  {volume} {113}},\ \bibinfo {pages} {053603} (\bibinfo {year}
  {2014})}\BibitemShut {NoStop}%
\bibitem [{\citenamefont {Wang}\ \emph {et~al.}(2018)\citenamefont {Wang},
  \citenamefont {Paesani}, \citenamefont {Ding}, \citenamefont {Santagati},
  \citenamefont {Skrzypczyk}, \citenamefont {Salavrakos}, \citenamefont {Tura},
  \citenamefont {Augusiak}, \citenamefont {Man{\v{c}}inska}, \citenamefont
  {Bacco} \emph {et~al.}}]{wang2018multidimensional}%
  \BibitemOpen
  \bibfield  {author} {\bibinfo {author} {\bibfnamefont {J.}~\bibnamefont
  {Wang}}, \bibinfo {author} {\bibfnamefont {S.}~\bibnamefont {Paesani}},
  \bibinfo {author} {\bibfnamefont {Y.}~\bibnamefont {Ding}}, \bibinfo {author}
  {\bibfnamefont {R.}~\bibnamefont {Santagati}}, \bibinfo {author}
  {\bibfnamefont {P.}~\bibnamefont {Skrzypczyk}}, \bibinfo {author}
  {\bibfnamefont {A.}~\bibnamefont {Salavrakos}}, \bibinfo {author}
  {\bibfnamefont {J.}~\bibnamefont {Tura}}, \bibinfo {author} {\bibfnamefont
  {R.}~\bibnamefont {Augusiak}}, \bibinfo {author} {\bibfnamefont
  {L.}~\bibnamefont {Man{\v{c}}inska}}, \bibinfo {author} {\bibfnamefont
  {D.}~\bibnamefont {Bacco}},  \emph {et~al.},\ }\href@noop {} {\bibfield
  {journal} {\bibinfo  {journal} {Science}\ }\textbf {\bibinfo {volume}
  {360}},\ \bibinfo {pages} {285} (\bibinfo {year} {2018})}\BibitemShut
  {NoStop}%
\bibitem [{\citenamefont {Cerf}\ \emph {et~al.}(2002)\citenamefont {Cerf},
  \citenamefont {Bourennane}, \citenamefont {Karlsson},\ and\ \citenamefont
  {Gisin}}]{cerf2002security}%
  \BibitemOpen
  \bibfield  {author} {\bibinfo {author} {\bibfnamefont {N.~J.}\ \bibnamefont
  {Cerf}}, \bibinfo {author} {\bibfnamefont {M.}~\bibnamefont {Bourennane}},
  \bibinfo {author} {\bibfnamefont {A.}~\bibnamefont {Karlsson}}, \ and\
  \bibinfo {author} {\bibfnamefont {N.}~\bibnamefont {Gisin}},\ }\href@noop {}
  {\bibfield  {journal} {\bibinfo  {journal} {Physical Review Letters}\
  }\textbf {\bibinfo {volume} {88}},\ \bibinfo {pages} {127902} (\bibinfo
  {year} {2002})}\BibitemShut {NoStop}%
\bibitem [{\citenamefont {Bouchard}\ \emph {et~al.}(2017)\citenamefont
  {Bouchard}, \citenamefont {Fickler}, \citenamefont {Boyd},\ and\
  \citenamefont {Karimi}}]{bouchard2017high}%
  \BibitemOpen
  \bibfield  {author} {\bibinfo {author} {\bibfnamefont {F.}~\bibnamefont
  {Bouchard}}, \bibinfo {author} {\bibfnamefont {R.}~\bibnamefont {Fickler}},
  \bibinfo {author} {\bibfnamefont {R.~W.}\ \bibnamefont {Boyd}}, \ and\
  \bibinfo {author} {\bibfnamefont {E.}~\bibnamefont {Karimi}},\ }\href@noop {}
  {\bibfield  {journal} {\bibinfo  {journal} {Science Advances}\ }\textbf
  {\bibinfo {volume} {3}},\ \bibinfo {pages} {e1601915} (\bibinfo {year}
  {2017})}\BibitemShut {NoStop}%
\bibitem [{\citenamefont {Huang}\ \emph {et~al.}(2022)\citenamefont {Huang},
  \citenamefont {Joshi}, \citenamefont {Aktas}, \citenamefont {Lupo},
  \citenamefont {Quintavalle}, \citenamefont {Venkatachalam}, \citenamefont
  {Wengerowsky}, \citenamefont {Lon{\v{c}}ari{\'c}}, \citenamefont {Neumann},
  \citenamefont {Liu} \emph {et~al.}}]{huang2022experimental}%
  \BibitemOpen
  \bibfield  {author} {\bibinfo {author} {\bibfnamefont {Z.}~\bibnamefont
  {Huang}}, \bibinfo {author} {\bibfnamefont {S.~K.}\ \bibnamefont {Joshi}},
  \bibinfo {author} {\bibfnamefont {D.}~\bibnamefont {Aktas}}, \bibinfo
  {author} {\bibfnamefont {C.}~\bibnamefont {Lupo}}, \bibinfo {author}
  {\bibfnamefont {A.~O.}\ \bibnamefont {Quintavalle}}, \bibinfo {author}
  {\bibfnamefont {N.}~\bibnamefont {Venkatachalam}}, \bibinfo {author}
  {\bibfnamefont {S.}~\bibnamefont {Wengerowsky}}, \bibinfo {author}
  {\bibfnamefont {M.}~\bibnamefont {Lon{\v{c}}ari{\'c}}}, \bibinfo {author}
  {\bibfnamefont {S.~P.}\ \bibnamefont {Neumann}}, \bibinfo {author}
  {\bibfnamefont {B.}~\bibnamefont {Liu}},  \emph {et~al.},\ }\href@noop {}
  {\bibfield  {journal} {\bibinfo  {journal} {npj Quantum Information}\
  }\textbf {\bibinfo {volume} {8}},\ \bibinfo {pages} {1} (\bibinfo {year}
  {2022})}\BibitemShut {NoStop}%
\bibitem [{\citenamefont {Paesani}\ \emph {et~al.}(2021)\citenamefont
  {Paesani}, \citenamefont {Bulmer}, \citenamefont {Jones}, \citenamefont
  {Santagati},\ and\ \citenamefont {Laing}}]{paesani2021scheme}%
  \BibitemOpen
  \bibfield  {author} {\bibinfo {author} {\bibfnamefont {S.}~\bibnamefont
  {Paesani}}, \bibinfo {author} {\bibfnamefont {J.~F.}\ \bibnamefont {Bulmer}},
  \bibinfo {author} {\bibfnamefont {A.~E.}\ \bibnamefont {Jones}}, \bibinfo
  {author} {\bibfnamefont {R.}~\bibnamefont {Santagati}}, \ and\ \bibinfo
  {author} {\bibfnamefont {A.}~\bibnamefont {Laing}},\ }\href@noop {}
  {\bibfield  {journal} {\bibinfo  {journal} {Physical Review Letters}\
  }\textbf {\bibinfo {volume} {126}},\ \bibinfo {pages} {230504} (\bibinfo
  {year} {2021})}\BibitemShut {NoStop}%
\bibitem [{\citenamefont {Peters}\ \emph {et~al.}(2021)\citenamefont {Peters},
  \citenamefont {Caldeira}, \citenamefont {Ho}, \citenamefont {Leichenauer},
  \citenamefont {Mohseni}, \citenamefont {Neven}, \citenamefont {Spentzouris},
  \citenamefont {Strain},\ and\ \citenamefont {Perdue}}]{peters2021machine}%
  \BibitemOpen
  \bibfield  {author} {\bibinfo {author} {\bibfnamefont {E.}~\bibnamefont
  {Peters}}, \bibinfo {author} {\bibfnamefont {J.}~\bibnamefont {Caldeira}},
  \bibinfo {author} {\bibfnamefont {A.}~\bibnamefont {Ho}}, \bibinfo {author}
  {\bibfnamefont {S.}~\bibnamefont {Leichenauer}}, \bibinfo {author}
  {\bibfnamefont {M.}~\bibnamefont {Mohseni}}, \bibinfo {author} {\bibfnamefont
  {H.}~\bibnamefont {Neven}}, \bibinfo {author} {\bibfnamefont
  {P.}~\bibnamefont {Spentzouris}}, \bibinfo {author} {\bibfnamefont
  {D.}~\bibnamefont {Strain}}, \ and\ \bibinfo {author} {\bibfnamefont {G.~N.}\
  \bibnamefont {Perdue}},\ }\href@noop {} {\bibfield  {journal} {\bibinfo
  {journal} {npj Quantum Information}\ }\textbf {\bibinfo {volume} {7}},\
  \bibinfo {pages} {1} (\bibinfo {year} {2021})}\BibitemShut {NoStop}%
\bibitem [{\citenamefont {Li}\ \emph {et~al.}(2020)\citenamefont {Li},
  \citenamefont {Liu}, \citenamefont {Ren}, \citenamefont {Wang}, \citenamefont
  {Su}, \citenamefont {Chen}, \citenamefont {Chu}, \citenamefont {Kuo},
  \citenamefont {Liu}, \citenamefont {Zang} \emph {et~al.}}]{li2020metalens}%
  \BibitemOpen
  \bibfield  {author} {\bibinfo {author} {\bibfnamefont {L.}~\bibnamefont
  {Li}}, \bibinfo {author} {\bibfnamefont {Z.}~\bibnamefont {Liu}}, \bibinfo
  {author} {\bibfnamefont {X.}~\bibnamefont {Ren}}, \bibinfo {author}
  {\bibfnamefont {S.}~\bibnamefont {Wang}}, \bibinfo {author} {\bibfnamefont
  {V.-C.}\ \bibnamefont {Su}}, \bibinfo {author} {\bibfnamefont {M.-K.}\
  \bibnamefont {Chen}}, \bibinfo {author} {\bibfnamefont {C.~H.}\ \bibnamefont
  {Chu}}, \bibinfo {author} {\bibfnamefont {H.~Y.}\ \bibnamefont {Kuo}},
  \bibinfo {author} {\bibfnamefont {B.}~\bibnamefont {Liu}}, \bibinfo {author}
  {\bibfnamefont {W.}~\bibnamefont {Zang}},  \emph {et~al.},\ }\href@noop {}
  {\bibfield  {journal} {\bibinfo  {journal} {Science}\ }\textbf {\bibinfo
  {volume} {368}},\ \bibinfo {pages} {1487} (\bibinfo {year}
  {2020})}\BibitemShut {NoStop}%
\bibitem [{\citenamefont {Erhard}\ \emph {et~al.}(2020)\citenamefont {Erhard},
  \citenamefont {Krenn},\ and\ \citenamefont {Zeilinger}}]{erhard2020advances}%
  \BibitemOpen
  \bibfield  {author} {\bibinfo {author} {\bibfnamefont {M.}~\bibnamefont
  {Erhard}}, \bibinfo {author} {\bibfnamefont {M.}~\bibnamefont {Krenn}}, \
  and\ \bibinfo {author} {\bibfnamefont {A.}~\bibnamefont {Zeilinger}},\
  }\href@noop {} {\bibfield  {journal} {\bibinfo  {journal} {Nature Reviews
  Physics}\ }\textbf {\bibinfo {volume} {2}},\ \bibinfo {pages} {365} (\bibinfo
  {year} {2020})}\BibitemShut {NoStop}%
\bibitem [{\citenamefont {Caspani}\ \emph {et~al.}(2016)\citenamefont
  {Caspani}, \citenamefont {Reimer}, \citenamefont {Kues}, \citenamefont
  {Roztocki}, \citenamefont {Clerici}, \citenamefont {Wetzel}, \citenamefont
  {Jestin}, \citenamefont {Ferrera}, \citenamefont {Peccianti}, \citenamefont
  {Pasquazi} \emph {et~al.}}]{caspani2016}%
  \BibitemOpen
  \bibfield  {author} {\bibinfo {author} {\bibfnamefont {L.}~\bibnamefont
  {Caspani}}, \bibinfo {author} {\bibfnamefont {C.}~\bibnamefont {Reimer}},
  \bibinfo {author} {\bibfnamefont {M.}~\bibnamefont {Kues}}, \bibinfo {author}
  {\bibfnamefont {P.}~\bibnamefont {Roztocki}}, \bibinfo {author}
  {\bibfnamefont {M.}~\bibnamefont {Clerici}}, \bibinfo {author} {\bibfnamefont
  {B.}~\bibnamefont {Wetzel}}, \bibinfo {author} {\bibfnamefont
  {Y.}~\bibnamefont {Jestin}}, \bibinfo {author} {\bibfnamefont
  {M.}~\bibnamefont {Ferrera}}, \bibinfo {author} {\bibfnamefont
  {M.}~\bibnamefont {Peccianti}}, \bibinfo {author} {\bibfnamefont
  {A.}~\bibnamefont {Pasquazi}},  \emph {et~al.},\ }\href@noop {} {\bibfield
  {journal} {\bibinfo  {journal} {Nanophotonics}\ }\textbf {\bibinfo {volume}
  {5}},\ \bibinfo {pages} {351} (\bibinfo {year} {2016})}\BibitemShut {NoStop}%
\bibitem [{\citenamefont {Chang}\ \emph {et~al.}(2021)\citenamefont {Chang},
  \citenamefont {Cheng}, \citenamefont {Sarihan}, \citenamefont {Vinod},
  \citenamefont {Lee}, \citenamefont {Zhong}, \citenamefont {Gong},
  \citenamefont {Xie}, \citenamefont {Shapiro}, \citenamefont {Wong} \emph
  {et~al.}}]{chang2021}%
  \BibitemOpen
  \bibfield  {author} {\bibinfo {author} {\bibfnamefont {K.-C.}\ \bibnamefont
  {Chang}}, \bibinfo {author} {\bibfnamefont {X.}~\bibnamefont {Cheng}},
  \bibinfo {author} {\bibfnamefont {M.~C.}\ \bibnamefont {Sarihan}}, \bibinfo
  {author} {\bibfnamefont {A.~K.}\ \bibnamefont {Vinod}}, \bibinfo {author}
  {\bibfnamefont {Y.~S.}\ \bibnamefont {Lee}}, \bibinfo {author} {\bibfnamefont
  {T.}~\bibnamefont {Zhong}}, \bibinfo {author} {\bibfnamefont {Y.-X.}\
  \bibnamefont {Gong}}, \bibinfo {author} {\bibfnamefont {Z.}~\bibnamefont
  {Xie}}, \bibinfo {author} {\bibfnamefont {J.~H.}\ \bibnamefont {Shapiro}},
  \bibinfo {author} {\bibfnamefont {F.~N.}\ \bibnamefont {Wong}},  \emph
  {et~al.},\ }\href@noop {} {\bibfield  {journal} {\bibinfo  {journal} {npj
  Quantum Information}\ }\textbf {\bibinfo {volume} {7}},\ \bibinfo {pages} {1}
  (\bibinfo {year} {2021})}\BibitemShut {NoStop}%
\bibitem [{\citenamefont {Ma}\ \emph {et~al.}(2020)\citenamefont {Ma},
  \citenamefont {Chen}, \citenamefont {Li}, \citenamefont {Tang}, \citenamefont
  {Sua}, \citenamefont {Fan},\ and\ \citenamefont {Huang}}]{ma2020ultrabright}%
  \BibitemOpen
  \bibfield  {author} {\bibinfo {author} {\bibfnamefont {Z.}~\bibnamefont
  {Ma}}, \bibinfo {author} {\bibfnamefont {J.-Y.}\ \bibnamefont {Chen}},
  \bibinfo {author} {\bibfnamefont {Z.}~\bibnamefont {Li}}, \bibinfo {author}
  {\bibfnamefont {C.}~\bibnamefont {Tang}}, \bibinfo {author} {\bibfnamefont
  {Y.~M.}\ \bibnamefont {Sua}}, \bibinfo {author} {\bibfnamefont
  {H.}~\bibnamefont {Fan}}, \ and\ \bibinfo {author} {\bibfnamefont {Y.-P.}\
  \bibnamefont {Huang}},\ }\href@noop {} {\bibfield  {journal} {\bibinfo
  {journal} {Physical Review Letters}\ }\textbf {\bibinfo {volume} {125}},\
  \bibinfo {pages} {263602} (\bibinfo {year} {2020})}\BibitemShut {NoStop}%
\bibitem [{\citenamefont {Luo}\ \emph {et~al.}(2017)\citenamefont {Luo},
  \citenamefont {Jiang}, \citenamefont {Rogers}, \citenamefont {Liang},
  \citenamefont {He},\ and\ \citenamefont {Lin}}]{luo2017chip}%
  \BibitemOpen
  \bibfield  {author} {\bibinfo {author} {\bibfnamefont {R.}~\bibnamefont
  {Luo}}, \bibinfo {author} {\bibfnamefont {H.}~\bibnamefont {Jiang}}, \bibinfo
  {author} {\bibfnamefont {S.}~\bibnamefont {Rogers}}, \bibinfo {author}
  {\bibfnamefont {H.}~\bibnamefont {Liang}}, \bibinfo {author} {\bibfnamefont
  {Y.}~\bibnamefont {He}}, \ and\ \bibinfo {author} {\bibfnamefont
  {Q.}~\bibnamefont {Lin}},\ }\href@noop {} {\bibfield  {journal} {\bibinfo
  {journal} {Optics Express}\ }\textbf {\bibinfo {volume} {25}},\ \bibinfo
  {pages} {24531} (\bibinfo {year} {2017})}\BibitemShut {NoStop}%
\bibitem [{\citenamefont {Xu}\ \emph {et~al.}(2021)\citenamefont {Xu},
  \citenamefont {Chen}, \citenamefont {Lin}, \citenamefont {Feng},
  \citenamefont {Niu}, \citenamefont {Zhou}, \citenamefont {Gao}, \citenamefont
  {Dong}, \citenamefont {Guo}, \citenamefont {Gong} \emph
  {et~al.}}]{xu2021spectrally}%
  \BibitemOpen
  \bibfield  {author} {\bibinfo {author} {\bibfnamefont {B.-Y.}\ \bibnamefont
  {Xu}}, \bibinfo {author} {\bibfnamefont {L.-K.}\ \bibnamefont {Chen}},
  \bibinfo {author} {\bibfnamefont {J.}~\bibnamefont {Lin}}, \bibinfo {author}
  {\bibfnamefont {L.-T.}\ \bibnamefont {Feng}}, \bibinfo {author}
  {\bibfnamefont {R.}~\bibnamefont {Niu}}, \bibinfo {author} {\bibfnamefont
  {Z.-y.}\ \bibnamefont {Zhou}}, \bibinfo {author} {\bibfnamefont
  {R.}~\bibnamefont {Gao}}, \bibinfo {author} {\bibfnamefont {C.-H.}\
  \bibnamefont {Dong}}, \bibinfo {author} {\bibfnamefont {G.-C.}\ \bibnamefont
  {Guo}}, \bibinfo {author} {\bibfnamefont {Q.}~\bibnamefont {Gong}},  \emph
  {et~al.},\ }\href@noop {} {\bibfield  {journal} {\bibinfo  {journal} {arXiv
  preprint arXiv:2110.08997}\ } (\bibinfo {year} {2021})}\BibitemShut {NoStop}%
\bibitem [{\citenamefont {Guo}\ \emph {et~al.}(2017)\citenamefont {Guo},
  \citenamefont {Zou}, \citenamefont {Schuck}, \citenamefont {Jung},
  \citenamefont {Cheng},\ and\ \citenamefont {Tang}}]{guo2017parametric}%
  \BibitemOpen
  \bibfield  {author} {\bibinfo {author} {\bibfnamefont {X.}~\bibnamefont
  {Guo}}, \bibinfo {author} {\bibfnamefont {C.-l.}\ \bibnamefont {Zou}},
  \bibinfo {author} {\bibfnamefont {C.}~\bibnamefont {Schuck}}, \bibinfo
  {author} {\bibfnamefont {H.}~\bibnamefont {Jung}}, \bibinfo {author}
  {\bibfnamefont {R.}~\bibnamefont {Cheng}}, \ and\ \bibinfo {author}
  {\bibfnamefont {H.~X.}\ \bibnamefont {Tang}},\ }\href@noop {} {\bibfield
  {journal} {\bibinfo  {journal} {Light: Science \& Applications}\ }\textbf
  {\bibinfo {volume} {6}},\ \bibinfo {pages} {e16249} (\bibinfo {year}
  {2017})}\BibitemShut {NoStop}%
\bibitem [{\citenamefont {Ong}\ \emph {et~al.}(2011)\citenamefont {Ong},
  \citenamefont {Cooper}, \citenamefont {Gupta}, \citenamefont {Green},
  \citenamefont {Assefa}, \citenamefont {Xia},\ and\ \citenamefont
  {Mookherjea}}]{ongsilicon2011}%
  \BibitemOpen
  \bibfield  {author} {\bibinfo {author} {\bibfnamefont {J.~R.}\ \bibnamefont
  {Ong}}, \bibinfo {author} {\bibfnamefont {M.~L.}\ \bibnamefont {Cooper}},
  \bibinfo {author} {\bibfnamefont {G.}~\bibnamefont {Gupta}}, \bibinfo
  {author} {\bibfnamefont {W.~M.}\ \bibnamefont {Green}}, \bibinfo {author}
  {\bibfnamefont {S.}~\bibnamefont {Assefa}}, \bibinfo {author} {\bibfnamefont
  {F.}~\bibnamefont {Xia}}, \ and\ \bibinfo {author} {\bibfnamefont
  {S.}~\bibnamefont {Mookherjea}},\ }\href@noop {} {\bibfield  {journal}
  {\bibinfo  {journal} {Optics Letters}\ }\textbf {\bibinfo {volume} {36}},\
  \bibinfo {pages} {2964} (\bibinfo {year} {2011})}\BibitemShut {NoStop}%
\bibitem [{\citenamefont {Kumar}\ \emph {et~al.}(2013)\citenamefont {Kumar},
  \citenamefont {Ong}, \citenamefont {Recchio}, \citenamefont {Srinivasan},\
  and\ \citenamefont {Mookherjea}}]{kumarsilicon2013}%
  \BibitemOpen
  \bibfield  {author} {\bibinfo {author} {\bibfnamefont {R.}~\bibnamefont
  {Kumar}}, \bibinfo {author} {\bibfnamefont {J.~R.}\ \bibnamefont {Ong}},
  \bibinfo {author} {\bibfnamefont {J.}~\bibnamefont {Recchio}}, \bibinfo
  {author} {\bibfnamefont {K.}~\bibnamefont {Srinivasan}}, \ and\ \bibinfo
  {author} {\bibfnamefont {S.}~\bibnamefont {Mookherjea}},\ }\href@noop {}
  {\bibfield  {journal} {\bibinfo  {journal} {Optics Letters}\ }\textbf
  {\bibinfo {volume} {38}},\ \bibinfo {pages} {2969} (\bibinfo {year}
  {2013})}\BibitemShut {NoStop}%
\bibitem [{\citenamefont {Kumar}\ \emph {et~al.}(2015)\citenamefont {Kumar},
  \citenamefont {Savanier}, \citenamefont {Ong},\ and\ \citenamefont
  {Mookherjea}}]{kumarsilicon2015}%
  \BibitemOpen
  \bibfield  {author} {\bibinfo {author} {\bibfnamefont {R.}~\bibnamefont
  {Kumar}}, \bibinfo {author} {\bibfnamefont {M.}~\bibnamefont {Savanier}},
  \bibinfo {author} {\bibfnamefont {J.~R.}\ \bibnamefont {Ong}}, \ and\
  \bibinfo {author} {\bibfnamefont {S.}~\bibnamefont {Mookherjea}},\
  }\href@noop {} {\bibfield  {journal} {\bibinfo  {journal} {Optics Express}\
  }\textbf {\bibinfo {volume} {23}},\ \bibinfo {pages} {19318} (\bibinfo {year}
  {2015})}\BibitemShut {NoStop}%
\bibitem [{\citenamefont {Jiang}\ \emph {et~al.}(2015)\citenamefont {Jiang},
  \citenamefont {Lu}, \citenamefont {Zhang}, \citenamefont {Painter},\ and\
  \citenamefont {Lin}}]{jiang2015silicon}%
  \BibitemOpen
  \bibfield  {author} {\bibinfo {author} {\bibfnamefont {W.~C.}\ \bibnamefont
  {Jiang}}, \bibinfo {author} {\bibfnamefont {X.}~\bibnamefont {Lu}}, \bibinfo
  {author} {\bibfnamefont {J.}~\bibnamefont {Zhang}}, \bibinfo {author}
  {\bibfnamefont {O.}~\bibnamefont {Painter}}, \ and\ \bibinfo {author}
  {\bibfnamefont {Q.}~\bibnamefont {Lin}},\ }\href@noop {} {\bibfield
  {journal} {\bibinfo  {journal} {Optics Express}\ }\textbf {\bibinfo {volume}
  {23}},\ \bibinfo {pages} {20884} (\bibinfo {year} {2015})}\BibitemShut
  {NoStop}%
\bibitem [{\citenamefont {Ramelow}\ \emph {et~al.}(2015)\citenamefont
  {Ramelow}, \citenamefont {Farsi}, \citenamefont {Clemmen}, \citenamefont
  {Orquiza}, \citenamefont {Luke}, \citenamefont {Lipson},\ and\ \citenamefont
  {Gaeta}}]{ramelow2015}%
  \BibitemOpen
  \bibfield  {author} {\bibinfo {author} {\bibfnamefont {S.}~\bibnamefont
  {Ramelow}}, \bibinfo {author} {\bibfnamefont {A.}~\bibnamefont {Farsi}},
  \bibinfo {author} {\bibfnamefont {S.}~\bibnamefont {Clemmen}}, \bibinfo
  {author} {\bibfnamefont {D.}~\bibnamefont {Orquiza}}, \bibinfo {author}
  {\bibfnamefont {K.}~\bibnamefont {Luke}}, \bibinfo {author} {\bibfnamefont
  {M.}~\bibnamefont {Lipson}}, \ and\ \bibinfo {author} {\bibfnamefont {A.~L.}\
  \bibnamefont {Gaeta}},\ }\href@noop {} {\bibfield  {journal} {\bibinfo
  {journal} {arXiv preprint arXiv:1508.04358}\ } (\bibinfo {year}
  {2015})}\BibitemShut {NoStop}%
\bibitem [{\citenamefont {Jaramillo-Villegas}\ \emph
  {et~al.}(2017)\citenamefont {Jaramillo-Villegas}, \citenamefont {Imany},
  \citenamefont {Odele}, \citenamefont {Leaird}, \citenamefont {Ou},
  \citenamefont {Qi},\ and\ \citenamefont {Weiner}}]{jaramillo2017}%
  \BibitemOpen
  \bibfield  {author} {\bibinfo {author} {\bibfnamefont {J.~A.}\ \bibnamefont
  {Jaramillo-Villegas}}, \bibinfo {author} {\bibfnamefont {P.}~\bibnamefont
  {Imany}}, \bibinfo {author} {\bibfnamefont {O.~D.}\ \bibnamefont {Odele}},
  \bibinfo {author} {\bibfnamefont {D.~E.}\ \bibnamefont {Leaird}}, \bibinfo
  {author} {\bibfnamefont {Z.-Y.}\ \bibnamefont {Ou}}, \bibinfo {author}
  {\bibfnamefont {M.}~\bibnamefont {Qi}}, \ and\ \bibinfo {author}
  {\bibfnamefont {A.~M.}\ \bibnamefont {Weiner}},\ }\href@noop {} {\bibfield
  {journal} {\bibinfo  {journal} {Optica}\ }\textbf {\bibinfo {volume} {4}},\
  \bibinfo {pages} {655} (\bibinfo {year} {2017})}\BibitemShut {NoStop}%
\bibitem [{\citenamefont {Imany}\ \emph {et~al.}(2018)\citenamefont {Imany},
  \citenamefont {Jaramillo-Villegas}, \citenamefont {Odele}, \citenamefont
  {Han}, \citenamefont {Leaird}, \citenamefont {Lukens}, \citenamefont
  {Lougovski}, \citenamefont {Qi},\ and\ \citenamefont {Weiner}}]{imany2018}%
  \BibitemOpen
  \bibfield  {author} {\bibinfo {author} {\bibfnamefont {P.}~\bibnamefont
  {Imany}}, \bibinfo {author} {\bibfnamefont {J.~A.}\ \bibnamefont
  {Jaramillo-Villegas}}, \bibinfo {author} {\bibfnamefont {O.~D.}\ \bibnamefont
  {Odele}}, \bibinfo {author} {\bibfnamefont {K.}~\bibnamefont {Han}}, \bibinfo
  {author} {\bibfnamefont {D.~E.}\ \bibnamefont {Leaird}}, \bibinfo {author}
  {\bibfnamefont {J.~M.}\ \bibnamefont {Lukens}}, \bibinfo {author}
  {\bibfnamefont {P.}~\bibnamefont {Lougovski}}, \bibinfo {author}
  {\bibfnamefont {M.}~\bibnamefont {Qi}}, \ and\ \bibinfo {author}
  {\bibfnamefont {A.~M.}\ \bibnamefont {Weiner}},\ }\href@noop {} {\bibfield
  {journal} {\bibinfo  {journal} {Optics Express}\ }\textbf {\bibinfo {volume}
  {26}},\ \bibinfo {pages} {1825} (\bibinfo {year} {2018})}\BibitemShut
  {NoStop}%
\bibitem [{\citenamefont {Lu}\ \emph {et~al.}(2019)\citenamefont {Lu},
  \citenamefont {Li}, \citenamefont {Westly}, \citenamefont {Moille},
  \citenamefont {Singh}, \citenamefont {Anant},\ and\ \citenamefont
  {Srinivasan}}]{lu2019}%
  \BibitemOpen
  \bibfield  {author} {\bibinfo {author} {\bibfnamefont {X.}~\bibnamefont
  {Lu}}, \bibinfo {author} {\bibfnamefont {Q.}~\bibnamefont {Li}}, \bibinfo
  {author} {\bibfnamefont {D.~A.}\ \bibnamefont {Westly}}, \bibinfo {author}
  {\bibfnamefont {G.}~\bibnamefont {Moille}}, \bibinfo {author} {\bibfnamefont
  {A.}~\bibnamefont {Singh}}, \bibinfo {author} {\bibfnamefont
  {V.}~\bibnamefont {Anant}}, \ and\ \bibinfo {author} {\bibfnamefont
  {K.}~\bibnamefont {Srinivasan}},\ }\href@noop {} {\bibfield  {journal}
  {\bibinfo  {journal} {Nature Physics}\ }\textbf {\bibinfo {volume} {15}},\
  \bibinfo {pages} {373} (\bibinfo {year} {2019})}\BibitemShut {NoStop}%
\bibitem [{\citenamefont {Samara}\ \emph {et~al.}(2019)\citenamefont {Samara},
  \citenamefont {Martin}, \citenamefont {Autebert}, \citenamefont {Karpov},
  \citenamefont {Kippenberg}, \citenamefont {Zbinden},\ and\ \citenamefont
  {Thew}}]{samara2019}%
  \BibitemOpen
  \bibfield  {author} {\bibinfo {author} {\bibfnamefont {F.}~\bibnamefont
  {Samara}}, \bibinfo {author} {\bibfnamefont {A.}~\bibnamefont {Martin}},
  \bibinfo {author} {\bibfnamefont {C.}~\bibnamefont {Autebert}}, \bibinfo
  {author} {\bibfnamefont {M.}~\bibnamefont {Karpov}}, \bibinfo {author}
  {\bibfnamefont {T.~J.}\ \bibnamefont {Kippenberg}}, \bibinfo {author}
  {\bibfnamefont {H.}~\bibnamefont {Zbinden}}, \ and\ \bibinfo {author}
  {\bibfnamefont {R.}~\bibnamefont {Thew}},\ }\href@noop {} {\bibfield
  {journal} {\bibinfo  {journal} {Optics Express}\ }\textbf {\bibinfo {volume}
  {27}},\ \bibinfo {pages} {19309} (\bibinfo {year} {2019})}\BibitemShut
  {NoStop}%
\bibitem [{\citenamefont {Li}\ \emph {et~al.}(2019)\citenamefont {Li},
  \citenamefont {Singh}, \citenamefont {Lu}, \citenamefont {Lawall},
  \citenamefont {Verma}, \citenamefont {Mirin}, \citenamefont {Nam},\ and\
  \citenamefont {Srinivasan}}]{li2019tunable}%
  \BibitemOpen
  \bibfield  {author} {\bibinfo {author} {\bibfnamefont {Q.}~\bibnamefont
  {Li}}, \bibinfo {author} {\bibfnamefont {A.}~\bibnamefont {Singh}}, \bibinfo
  {author} {\bibfnamefont {X.}~\bibnamefont {Lu}}, \bibinfo {author}
  {\bibfnamefont {J.}~\bibnamefont {Lawall}}, \bibinfo {author} {\bibfnamefont
  {V.}~\bibnamefont {Verma}}, \bibinfo {author} {\bibfnamefont
  {R.}~\bibnamefont {Mirin}}, \bibinfo {author} {\bibfnamefont {S.~W.}\
  \bibnamefont {Nam}}, \ and\ \bibinfo {author} {\bibfnamefont
  {K.}~\bibnamefont {Srinivasan}},\ }\href@noop {} {\bibfield  {journal}
  {\bibinfo  {journal} {Physical Review Applied}\ }\textbf {\bibinfo {volume}
  {12}},\ \bibinfo {pages} {054054} (\bibinfo {year} {2019})}\BibitemShut
  {NoStop}%
\bibitem [{\citenamefont {Lu}\ \emph {et~al.}(2022{\natexlab{a}})\citenamefont
  {Lu}, \citenamefont {Myilswamy}, \citenamefont {Bennink}, \citenamefont
  {Seshadri}, \citenamefont {Alshaykh}, \citenamefont {Liu}, \citenamefont
  {Kippenberg}, \citenamefont {Leaird}, \citenamefont {Weiner},\ and\
  \citenamefont {Lukens}}]{lu2022bayesian}%
  \BibitemOpen
  \bibfield  {author} {\bibinfo {author} {\bibfnamefont {H.-H.}\ \bibnamefont
  {Lu}}, \bibinfo {author} {\bibfnamefont {K.~V.}\ \bibnamefont {Myilswamy}},
  \bibinfo {author} {\bibfnamefont {R.~S.}\ \bibnamefont {Bennink}}, \bibinfo
  {author} {\bibfnamefont {S.}~\bibnamefont {Seshadri}}, \bibinfo {author}
  {\bibfnamefont {M.~S.}\ \bibnamefont {Alshaykh}}, \bibinfo {author}
  {\bibfnamefont {J.}~\bibnamefont {Liu}}, \bibinfo {author} {\bibfnamefont
  {T.~J.}\ \bibnamefont {Kippenberg}}, \bibinfo {author} {\bibfnamefont
  {D.~E.}\ \bibnamefont {Leaird}}, \bibinfo {author} {\bibfnamefont {A.~M.}\
  \bibnamefont {Weiner}}, \ and\ \bibinfo {author} {\bibfnamefont {J.~M.}\
  \bibnamefont {Lukens}},\ }\href@noop {} {\bibfield  {journal} {\bibinfo
  {journal} {Nature Communications}\ }\textbf {\bibinfo {volume} {13}},\
  \bibinfo {pages} {1} (\bibinfo {year} {2022}{\natexlab{a}})}\BibitemShut
  {NoStop}%
\bibitem [{\citenamefont {Yin}\ \emph {et~al.}(2021)\citenamefont {Yin},
  \citenamefont {Sugiura}, \citenamefont {Takashima}, \citenamefont {Okamoto},
  \citenamefont {Qiu}, \citenamefont {Yokoyama},\ and\ \citenamefont
  {Takeuchi}}]{yin2021}%
  \BibitemOpen
  \bibfield  {author} {\bibinfo {author} {\bibfnamefont {Z.}~\bibnamefont
  {Yin}}, \bibinfo {author} {\bibfnamefont {K.}~\bibnamefont {Sugiura}},
  \bibinfo {author} {\bibfnamefont {H.}~\bibnamefont {Takashima}}, \bibinfo
  {author} {\bibfnamefont {R.}~\bibnamefont {Okamoto}}, \bibinfo {author}
  {\bibfnamefont {F.}~\bibnamefont {Qiu}}, \bibinfo {author} {\bibfnamefont
  {S.}~\bibnamefont {Yokoyama}}, \ and\ \bibinfo {author} {\bibfnamefont
  {S.}~\bibnamefont {Takeuchi}},\ }\href@noop {} {\bibfield  {journal}
  {\bibinfo  {journal} {Optics Express}\ }\textbf {\bibinfo {volume} {29}},\
  \bibinfo {pages} {4821} (\bibinfo {year} {2021})}\BibitemShut {NoStop}%
\bibitem [{\citenamefont {Samara}\ \emph {et~al.}(2021)\citenamefont {Samara},
  \citenamefont {Maring}, \citenamefont {Martin}, \citenamefont {Raja},
  \citenamefont {Kippenberg}, \citenamefont {Zbinden},\ and\ \citenamefont
  {Thew}}]{samara2021entanglement}%
  \BibitemOpen
  \bibfield  {author} {\bibinfo {author} {\bibfnamefont {F.}~\bibnamefont
  {Samara}}, \bibinfo {author} {\bibfnamefont {N.}~\bibnamefont {Maring}},
  \bibinfo {author} {\bibfnamefont {A.}~\bibnamefont {Martin}}, \bibinfo
  {author} {\bibfnamefont {A.~S.}\ \bibnamefont {Raja}}, \bibinfo {author}
  {\bibfnamefont {T.~J.}\ \bibnamefont {Kippenberg}}, \bibinfo {author}
  {\bibfnamefont {H.}~\bibnamefont {Zbinden}}, \ and\ \bibinfo {author}
  {\bibfnamefont {R.}~\bibnamefont {Thew}},\ }\href@noop {} {\bibfield
  {journal} {\bibinfo  {journal} {Quantum Science and Technology}\ }\textbf
  {\bibinfo {volume} {6}},\ \bibinfo {pages} {045024} (\bibinfo {year}
  {2021})}\BibitemShut {NoStop}%
\bibitem [{\citenamefont {Reimer}\ \emph {et~al.}(2014)\citenamefont {Reimer},
  \citenamefont {Caspani}, \citenamefont {Clerici}, \citenamefont {Ferrera},
  \citenamefont {Kues}, \citenamefont {Peccianti}, \citenamefont {Pasquazi},
  \citenamefont {Razzari}, \citenamefont {Little}, \citenamefont {Chu} \emph
  {et~al.}}]{reimer2014}%
  \BibitemOpen
  \bibfield  {author} {\bibinfo {author} {\bibfnamefont {C.}~\bibnamefont
  {Reimer}}, \bibinfo {author} {\bibfnamefont {L.}~\bibnamefont {Caspani}},
  \bibinfo {author} {\bibfnamefont {M.}~\bibnamefont {Clerici}}, \bibinfo
  {author} {\bibfnamefont {M.}~\bibnamefont {Ferrera}}, \bibinfo {author}
  {\bibfnamefont {M.}~\bibnamefont {Kues}}, \bibinfo {author} {\bibfnamefont
  {M.}~\bibnamefont {Peccianti}}, \bibinfo {author} {\bibfnamefont
  {A.}~\bibnamefont {Pasquazi}}, \bibinfo {author} {\bibfnamefont
  {L.}~\bibnamefont {Razzari}}, \bibinfo {author} {\bibfnamefont {B.~E.}\
  \bibnamefont {Little}}, \bibinfo {author} {\bibfnamefont {S.~T.}\
  \bibnamefont {Chu}},  \emph {et~al.},\ }\href@noop {} {\bibfield  {journal}
  {\bibinfo  {journal} {Optics Express}\ }\textbf {\bibinfo {volume} {22}},\
  \bibinfo {pages} {6535} (\bibinfo {year} {2014})}\BibitemShut {NoStop}%
\bibitem [{\citenamefont {Reimer}\ \emph {et~al.}(2016)\citenamefont {Reimer},
  \citenamefont {Kues}, \citenamefont {Roztocki}, \citenamefont {Wetzel},
  \citenamefont {Grazioso}, \citenamefont {Little}, \citenamefont {Chu},
  \citenamefont {Johnston}, \citenamefont {Bromberg}, \citenamefont {Caspani}
  \emph {et~al.}}]{reimer2016science}%
  \BibitemOpen
  \bibfield  {author} {\bibinfo {author} {\bibfnamefont {C.}~\bibnamefont
  {Reimer}}, \bibinfo {author} {\bibfnamefont {M.}~\bibnamefont {Kues}},
  \bibinfo {author} {\bibfnamefont {P.}~\bibnamefont {Roztocki}}, \bibinfo
  {author} {\bibfnamefont {B.}~\bibnamefont {Wetzel}}, \bibinfo {author}
  {\bibfnamefont {F.}~\bibnamefont {Grazioso}}, \bibinfo {author}
  {\bibfnamefont {B.~E.}\ \bibnamefont {Little}}, \bibinfo {author}
  {\bibfnamefont {S.~T.}\ \bibnamefont {Chu}}, \bibinfo {author} {\bibfnamefont
  {T.}~\bibnamefont {Johnston}}, \bibinfo {author} {\bibfnamefont
  {Y.}~\bibnamefont {Bromberg}}, \bibinfo {author} {\bibfnamefont
  {L.}~\bibnamefont {Caspani}},  \emph {et~al.},\ }\href@noop {} {\bibfield
  {journal} {\bibinfo  {journal} {Science}\ }\textbf {\bibinfo {volume}
  {351}},\ \bibinfo {pages} {1176} (\bibinfo {year} {2016})}\BibitemShut
  {NoStop}%
\bibitem [{\citenamefont {Reimer}\ \emph {et~al.}(2019)\citenamefont {Reimer},
  \citenamefont {Sciara}, \citenamefont {Roztocki}, \citenamefont {Islam},
  \citenamefont {Cort{\'e}s}, \citenamefont {Zhang}, \citenamefont {Fischer},
  \citenamefont {Loranger}, \citenamefont {Kashyap}, \citenamefont {Cino} \emph
  {et~al.}}]{reimer2019}%
  \BibitemOpen
  \bibfield  {author} {\bibinfo {author} {\bibfnamefont {C.}~\bibnamefont
  {Reimer}}, \bibinfo {author} {\bibfnamefont {S.}~\bibnamefont {Sciara}},
  \bibinfo {author} {\bibfnamefont {P.}~\bibnamefont {Roztocki}}, \bibinfo
  {author} {\bibfnamefont {M.}~\bibnamefont {Islam}}, \bibinfo {author}
  {\bibfnamefont {L.~R.}\ \bibnamefont {Cort{\'e}s}}, \bibinfo {author}
  {\bibfnamefont {Y.}~\bibnamefont {Zhang}}, \bibinfo {author} {\bibfnamefont
  {B.}~\bibnamefont {Fischer}}, \bibinfo {author} {\bibfnamefont
  {S.}~\bibnamefont {Loranger}}, \bibinfo {author} {\bibfnamefont
  {R.}~\bibnamefont {Kashyap}}, \bibinfo {author} {\bibfnamefont
  {A.}~\bibnamefont {Cino}},  \emph {et~al.},\ }\href@noop {} {\bibfield
  {journal} {\bibinfo  {journal} {Nature Physics}\ }\textbf {\bibinfo {volume}
  {15}},\ \bibinfo {pages} {148} (\bibinfo {year} {2019})}\BibitemShut
  {NoStop}%
\bibitem [{\citenamefont {Liu}\ \emph {et~al.}(2020)\citenamefont {Liu},
  \citenamefont {Weng}, \citenamefont {Afridi}, \citenamefont {Li},
  \citenamefont {Dai}, \citenamefont {Ma}, \citenamefont {Long}, \citenamefont
  {Zhang}, \citenamefont {Lu}, \citenamefont {Donegan} \emph
  {et~al.}}]{liu2020photolithography}%
  \BibitemOpen
  \bibfield  {author} {\bibinfo {author} {\bibfnamefont {J.}~\bibnamefont
  {Liu}}, \bibinfo {author} {\bibfnamefont {H.}~\bibnamefont {Weng}}, \bibinfo
  {author} {\bibfnamefont {A.~A.}\ \bibnamefont {Afridi}}, \bibinfo {author}
  {\bibfnamefont {J.}~\bibnamefont {Li}}, \bibinfo {author} {\bibfnamefont
  {J.}~\bibnamefont {Dai}}, \bibinfo {author} {\bibfnamefont {X.}~\bibnamefont
  {Ma}}, \bibinfo {author} {\bibfnamefont {H.}~\bibnamefont {Long}}, \bibinfo
  {author} {\bibfnamefont {Y.}~\bibnamefont {Zhang}}, \bibinfo {author}
  {\bibfnamefont {Q.}~\bibnamefont {Lu}}, \bibinfo {author} {\bibfnamefont
  {J.~F.}\ \bibnamefont {Donegan}},  \emph {et~al.},\ }\href@noop {} {\bibfield
   {journal} {\bibinfo  {journal} {Optics Express}\ }\textbf {\bibinfo {volume}
  {28}},\ \bibinfo {pages} {19270} (\bibinfo {year} {2020})}\BibitemShut
  {NoStop}%
\bibitem [{\citenamefont {Brown}\ and\ \citenamefont
  {Twiss}(1956)}]{brown1956}%
  \BibitemOpen
  \bibfield  {author} {\bibinfo {author} {\bibfnamefont {R.~H.}\ \bibnamefont
  {Brown}}\ and\ \bibinfo {author} {\bibfnamefont {R.~Q.}\ \bibnamefont
  {Twiss}},\ }\href@noop {} {\bibfield  {journal} {\bibinfo  {journal}
  {Nature}\ }\textbf {\bibinfo {volume} {177}},\ \bibinfo {pages} {27}
  (\bibinfo {year} {1956})}\BibitemShut {NoStop}%
\bibitem [{\citenamefont {Steudle}\ \emph {et~al.}(2012)\citenamefont
  {Steudle}, \citenamefont {Schietinger}, \citenamefont {H{\"o}ckel},
  \citenamefont {Dorenbos}, \citenamefont {Zadeh}, \citenamefont {Zwiller},\
  and\ \citenamefont {Benson}}]{steudle2012}%
  \BibitemOpen
  \bibfield  {author} {\bibinfo {author} {\bibfnamefont {G.~A.}\ \bibnamefont
  {Steudle}}, \bibinfo {author} {\bibfnamefont {S.}~\bibnamefont
  {Schietinger}}, \bibinfo {author} {\bibfnamefont {D.}~\bibnamefont
  {H{\"o}ckel}}, \bibinfo {author} {\bibfnamefont {S.~N.}\ \bibnamefont
  {Dorenbos}}, \bibinfo {author} {\bibfnamefont {I.~E.}\ \bibnamefont {Zadeh}},
  \bibinfo {author} {\bibfnamefont {V.}~\bibnamefont {Zwiller}}, \ and\
  \bibinfo {author} {\bibfnamefont {O.}~\bibnamefont {Benson}},\ }\href@noop {}
  {\bibfield  {journal} {\bibinfo  {journal} {Physical Review A}\ }\textbf
  {\bibinfo {volume} {86}},\ \bibinfo {pages} {053814} (\bibinfo {year}
  {2012})}\BibitemShut {NoStop}%
\bibitem [{\citenamefont {Franson}(1989)}]{franson1989bell}%
  \BibitemOpen
  \bibfield  {author} {\bibinfo {author} {\bibfnamefont {J.~D.}\ \bibnamefont
  {Franson}},\ }\href@noop {} {\bibfield  {journal} {\bibinfo  {journal}
  {Physical Review Letters}\ }\textbf {\bibinfo {volume} {62}},\ \bibinfo
  {pages} {2205} (\bibinfo {year} {1989})}\BibitemShut {NoStop}%
\bibitem [{\citenamefont {Kwiat}\ \emph {et~al.}(1993)\citenamefont {Kwiat},
  \citenamefont {Steinberg},\ and\ \citenamefont {Chiao}}]{kwiat1993high}%
  \BibitemOpen
  \bibfield  {author} {\bibinfo {author} {\bibfnamefont {P.~G.}\ \bibnamefont
  {Kwiat}}, \bibinfo {author} {\bibfnamefont {A.~M.}\ \bibnamefont
  {Steinberg}}, \ and\ \bibinfo {author} {\bibfnamefont {R.~Y.}\ \bibnamefont
  {Chiao}},\ }\href@noop {} {\bibfield  {journal} {\bibinfo  {journal}
  {Physical Review A}\ }\textbf {\bibinfo {volume} {47}},\ \bibinfo {pages}
  {R2472} (\bibinfo {year} {1993})}\BibitemShut {NoStop}%
\bibitem [{\citenamefont {Marcikic}\ \emph {et~al.}(2004)\citenamefont
  {Marcikic}, \citenamefont {De~Riedmatten}, \citenamefont {Tittel},
  \citenamefont {Zbinden}, \citenamefont {Legr{\'e}},\ and\ \citenamefont
  {Gisin}}]{marcikic2004}%
  \BibitemOpen
  \bibfield  {author} {\bibinfo {author} {\bibfnamefont {I.}~\bibnamefont
  {Marcikic}}, \bibinfo {author} {\bibfnamefont {H.}~\bibnamefont
  {De~Riedmatten}}, \bibinfo {author} {\bibfnamefont {W.}~\bibnamefont
  {Tittel}}, \bibinfo {author} {\bibfnamefont {H.}~\bibnamefont {Zbinden}},
  \bibinfo {author} {\bibfnamefont {M.}~\bibnamefont {Legr{\'e}}}, \ and\
  \bibinfo {author} {\bibfnamefont {N.}~\bibnamefont {Gisin}},\ }\href@noop {}
  {\bibfield  {journal} {\bibinfo  {journal} {Physical Review Letters}\
  }\textbf {\bibinfo {volume} {93}},\ \bibinfo {pages} {180502} (\bibinfo
  {year} {2004})}\BibitemShut {NoStop}%
\bibitem [{\citenamefont {Del’Haye}\ \emph {et~al.}(2007)\citenamefont
  {Del’Haye}, \citenamefont {Schliesser}, \citenamefont {Arcizet},
  \citenamefont {Wilken}, \citenamefont {Holzwarth},\ and\ \citenamefont
  {Kippenberg}}]{del2007optical}%
  \BibitemOpen
  \bibfield  {author} {\bibinfo {author} {\bibfnamefont {P.}~\bibnamefont
  {Del’Haye}}, \bibinfo {author} {\bibfnamefont {A.}~\bibnamefont
  {Schliesser}}, \bibinfo {author} {\bibfnamefont {O.}~\bibnamefont {Arcizet}},
  \bibinfo {author} {\bibfnamefont {T.}~\bibnamefont {Wilken}}, \bibinfo
  {author} {\bibfnamefont {R.}~\bibnamefont {Holzwarth}}, \ and\ \bibinfo
  {author} {\bibfnamefont {T.~J.}\ \bibnamefont {Kippenberg}},\ }\href@noop {}
  {\bibfield  {journal} {\bibinfo  {journal} {Nature}\ }\textbf {\bibinfo
  {volume} {450}},\ \bibinfo {pages} {1214} (\bibinfo {year}
  {2007})}\BibitemShut {NoStop}%
\bibitem [{\citenamefont {Kippenberg}\ \emph {et~al.}(2011)\citenamefont
  {Kippenberg}, \citenamefont {Holzwarth},\ and\ \citenamefont
  {Diddams}}]{kippenberg2011microresonator}%
  \BibitemOpen
  \bibfield  {author} {\bibinfo {author} {\bibfnamefont {T.~J.}\ \bibnamefont
  {Kippenberg}}, \bibinfo {author} {\bibfnamefont {R.}~\bibnamefont
  {Holzwarth}}, \ and\ \bibinfo {author} {\bibfnamefont {S.~A.}\ \bibnamefont
  {Diddams}},\ }\href@noop {} {\bibfield  {journal} {\bibinfo  {journal}
  {Science}\ }\textbf {\bibinfo {volume} {332}},\ \bibinfo {pages} {555}
  (\bibinfo {year} {2011})}\BibitemShut {NoStop}%
\bibitem [{\citenamefont {Herr}\ \emph {et~al.}(2012)\citenamefont {Herr},
  \citenamefont {Hartinger}, \citenamefont {Riemensberger}, \citenamefont
  {Wang}, \citenamefont {Gavartin}, \citenamefont {Holzwarth}, \citenamefont
  {Gorodetsky},\ and\ \citenamefont {Kippenberg}}]{herr2012universal}%
  \BibitemOpen
  \bibfield  {author} {\bibinfo {author} {\bibfnamefont {T.}~\bibnamefont
  {Herr}}, \bibinfo {author} {\bibfnamefont {K.}~\bibnamefont {Hartinger}},
  \bibinfo {author} {\bibfnamefont {J.}~\bibnamefont {Riemensberger}}, \bibinfo
  {author} {\bibfnamefont {C.}~\bibnamefont {Wang}}, \bibinfo {author}
  {\bibfnamefont {E.}~\bibnamefont {Gavartin}}, \bibinfo {author}
  {\bibfnamefont {R.}~\bibnamefont {Holzwarth}}, \bibinfo {author}
  {\bibfnamefont {M.}~\bibnamefont {Gorodetsky}}, \ and\ \bibinfo {author}
  {\bibfnamefont {T.}~\bibnamefont {Kippenberg}},\ }\href@noop {} {\bibfield
  {journal} {\bibinfo  {journal} {Nature photonics}\ }\textbf {\bibinfo
  {volume} {6}},\ \bibinfo {pages} {480} (\bibinfo {year} {2012})}\BibitemShut
  {NoStop}%
\bibitem [{\citenamefont {Zhou}\ \emph {et~al.}(2019)\citenamefont {Zhou},
  \citenamefont {Geng}, \citenamefont {Cui}, \citenamefont {Huang},
  \citenamefont {Zhou}, \citenamefont {Qiu},\ and\ \citenamefont
  {Wei~Wong}}]{zhou2019soliton}%
  \BibitemOpen
  \bibfield  {author} {\bibinfo {author} {\bibfnamefont {H.}~\bibnamefont
  {Zhou}}, \bibinfo {author} {\bibfnamefont {Y.}~\bibnamefont {Geng}}, \bibinfo
  {author} {\bibfnamefont {W.}~\bibnamefont {Cui}}, \bibinfo {author}
  {\bibfnamefont {S.-W.}\ \bibnamefont {Huang}}, \bibinfo {author}
  {\bibfnamefont {Q.}~\bibnamefont {Zhou}}, \bibinfo {author} {\bibfnamefont
  {K.}~\bibnamefont {Qiu}}, \ and\ \bibinfo {author} {\bibfnamefont
  {C.}~\bibnamefont {Wei~Wong}},\ }\href@noop {} {\bibfield  {journal}
  {\bibinfo  {journal} {Light: Science \& Applications}\ }\textbf {\bibinfo
  {volume} {8}},\ \bibinfo {pages} {1} (\bibinfo {year} {2019})}\BibitemShut
  {NoStop}%
\bibitem [{\citenamefont {Geng}\ \emph {et~al.}(2022)\citenamefont {Geng},
  \citenamefont {Zhou}, \citenamefont {Han}, \citenamefont {Cui}, \citenamefont
  {Zhang}, \citenamefont {Liu}, \citenamefont {Deng}, \citenamefont {Zhou},\
  and\ \citenamefont {Qiu}}]{geng2022coherent}%
  \BibitemOpen
  \bibfield  {author} {\bibinfo {author} {\bibfnamefont {Y.}~\bibnamefont
  {Geng}}, \bibinfo {author} {\bibfnamefont {H.}~\bibnamefont {Zhou}}, \bibinfo
  {author} {\bibfnamefont {X.}~\bibnamefont {Han}}, \bibinfo {author}
  {\bibfnamefont {W.}~\bibnamefont {Cui}}, \bibinfo {author} {\bibfnamefont
  {Q.}~\bibnamefont {Zhang}}, \bibinfo {author} {\bibfnamefont
  {B.}~\bibnamefont {Liu}}, \bibinfo {author} {\bibfnamefont {G.}~\bibnamefont
  {Deng}}, \bibinfo {author} {\bibfnamefont {Q.}~\bibnamefont {Zhou}}, \ and\
  \bibinfo {author} {\bibfnamefont {K.}~\bibnamefont {Qiu}},\ }\href@noop {}
  {\bibfield  {journal} {\bibinfo  {journal} {Nature Communications}\ }\textbf
  {\bibinfo {volume} {13}},\ \bibinfo {pages} {1} (\bibinfo {year}
  {2022})}\BibitemShut {NoStop}%
\bibitem [{\citenamefont {Wu}\ \emph {et~al.}(2020)\citenamefont {Wu},
  \citenamefont {Liu}, \citenamefont {Gu}, \citenamefont {Yu}, \citenamefont
  {Kong}, \citenamefont {Wang}, \citenamefont {Qiang}, \citenamefont {Wu},
  \citenamefont {Zhu}, \citenamefont {Yang} \emph {et~al.}}]{wu2020bright}%
  \BibitemOpen
  \bibfield  {author} {\bibinfo {author} {\bibfnamefont {C.}~\bibnamefont
  {Wu}}, \bibinfo {author} {\bibfnamefont {Y.}~\bibnamefont {Liu}}, \bibinfo
  {author} {\bibfnamefont {X.}~\bibnamefont {Gu}}, \bibinfo {author}
  {\bibfnamefont {X.}~\bibnamefont {Yu}}, \bibinfo {author} {\bibfnamefont
  {Y.}~\bibnamefont {Kong}}, \bibinfo {author} {\bibfnamefont {Y.}~\bibnamefont
  {Wang}}, \bibinfo {author} {\bibfnamefont {X.}~\bibnamefont {Qiang}},
  \bibinfo {author} {\bibfnamefont {J.}~\bibnamefont {Wu}}, \bibinfo {author}
  {\bibfnamefont {Z.}~\bibnamefont {Zhu}}, \bibinfo {author} {\bibfnamefont
  {X.}~\bibnamefont {Yang}},  \emph {et~al.},\ }\href@noop {} {\bibfield
  {journal} {\bibinfo  {journal} {Science China Physics, Mechanics \&
  Astronomy}\ }\textbf {\bibinfo {volume} {63}},\ \bibinfo {pages} {1}
  (\bibinfo {year} {2020})}\BibitemShut {NoStop}%
\bibitem [{\citenamefont {Peng}\ \emph {et~al.}(2014)\citenamefont {Peng},
  \citenamefont {{\"O}zdemir}, \citenamefont {Lei}, \citenamefont {Monifi},
  \citenamefont {Gianfreda}, \citenamefont {Long}, \citenamefont {Fan},
  \citenamefont {Nori}, \citenamefont {Bender},\ and\ \citenamefont
  {Yang}}]{peng2014parity}%
  \BibitemOpen
  \bibfield  {author} {\bibinfo {author} {\bibfnamefont {B.}~\bibnamefont
  {Peng}}, \bibinfo {author} {\bibfnamefont {{\c{S}}.~K.}\ \bibnamefont
  {{\"O}zdemir}}, \bibinfo {author} {\bibfnamefont {F.}~\bibnamefont {Lei}},
  \bibinfo {author} {\bibfnamefont {F.}~\bibnamefont {Monifi}}, \bibinfo
  {author} {\bibfnamefont {M.}~\bibnamefont {Gianfreda}}, \bibinfo {author}
  {\bibfnamefont {G.~L.}\ \bibnamefont {Long}}, \bibinfo {author}
  {\bibfnamefont {S.}~\bibnamefont {Fan}}, \bibinfo {author} {\bibfnamefont
  {F.}~\bibnamefont {Nori}}, \bibinfo {author} {\bibfnamefont {C.~M.}\
  \bibnamefont {Bender}}, \ and\ \bibinfo {author} {\bibfnamefont
  {L.}~\bibnamefont {Yang}},\ }\href@noop {} {\bibfield  {journal} {\bibinfo
  {journal} {Nature Physics}\ }\textbf {\bibinfo {volume} {10}},\ \bibinfo
  {pages} {394} (\bibinfo {year} {2014})}\BibitemShut {NoStop}%
\bibitem [{\citenamefont {Lu}\ \emph {et~al.}(2022{\natexlab{b}})\citenamefont
  {Lu}, \citenamefont {Chen}, \citenamefont {Zhang}, \citenamefont {Yang},
  \citenamefont {Chen}, \citenamefont {Zhang},\ and\ \citenamefont
  {Xu}}]{lu2022parity}%
  \BibitemOpen
  \bibfield  {author} {\bibinfo {author} {\bibfnamefont {X.}~\bibnamefont
  {Lu}}, \bibinfo {author} {\bibfnamefont {N.}~\bibnamefont {Chen}}, \bibinfo
  {author} {\bibfnamefont {B.}~\bibnamefont {Zhang}}, \bibinfo {author}
  {\bibfnamefont {H.}~\bibnamefont {Yang}}, \bibinfo {author} {\bibfnamefont
  {Y.}~\bibnamefont {Chen}}, \bibinfo {author} {\bibfnamefont {X.}~\bibnamefont
  {Zhang}}, \ and\ \bibinfo {author} {\bibfnamefont {J.}~\bibnamefont {Xu}},\
  }\href@noop {} {\bibfield  {journal} {\bibinfo  {journal} {Photonics}\
  }\textbf {\bibinfo {volume} {9}},\ \bibinfo {pages} {380} (\bibinfo {year}
  {2022}{\natexlab{b}})}\BibitemShut {NoStop}%
\bibitem [{\citenamefont {P{\'e}rez-Galacho}\ \emph {et~al.}(2017)\citenamefont
  {P{\'e}rez-Galacho}, \citenamefont {Alonso-Ramos}, \citenamefont {Mazeas},
  \citenamefont {Le~Roux}, \citenamefont {Oser}, \citenamefont {Zhang},
  \citenamefont {Marris-Morini}, \citenamefont {Labont{\'e}}, \citenamefont
  {Tanzilli}, \citenamefont {Cassan} \emph {et~al.}}]{perez2017optical}%
  \BibitemOpen
  \bibfield  {author} {\bibinfo {author} {\bibfnamefont {D.}~\bibnamefont
  {P{\'e}rez-Galacho}}, \bibinfo {author} {\bibfnamefont {C.}~\bibnamefont
  {Alonso-Ramos}}, \bibinfo {author} {\bibfnamefont {F.}~\bibnamefont
  {Mazeas}}, \bibinfo {author} {\bibfnamefont {X.}~\bibnamefont {Le~Roux}},
  \bibinfo {author} {\bibfnamefont {D.}~\bibnamefont {Oser}}, \bibinfo {author}
  {\bibfnamefont {W.}~\bibnamefont {Zhang}}, \bibinfo {author} {\bibfnamefont
  {D.}~\bibnamefont {Marris-Morini}}, \bibinfo {author} {\bibfnamefont
  {L.}~\bibnamefont {Labont{\'e}}}, \bibinfo {author} {\bibfnamefont
  {S.}~\bibnamefont {Tanzilli}}, \bibinfo {author} {\bibfnamefont
  {{\'E}.}~\bibnamefont {Cassan}},  \emph {et~al.},\ }\href@noop {} {\bibfield
  {journal} {\bibinfo  {journal} {Optics letters}\ }\textbf {\bibinfo {volume}
  {42}},\ \bibinfo {pages} {1468} (\bibinfo {year} {2017})}\BibitemShut
  {NoStop}%
\bibitem [{\citenamefont {Yu}\ \emph {et~al.}(2022)\citenamefont {Yu},
  \citenamefont {Yuan}, \citenamefont {Zhang}, \citenamefont {Zhang},
  \citenamefont {Li}, \citenamefont {Wang}, \citenamefont {Deng}, \citenamefont
  {You}, \citenamefont {Song}, \citenamefont {Wang} \emph
  {et~al.}}]{yu2022spectrally}%
  \BibitemOpen
  \bibfield  {author} {\bibinfo {author} {\bibfnamefont {H.}~\bibnamefont
  {Yu}}, \bibinfo {author} {\bibfnamefont {C.}~\bibnamefont {Yuan}}, \bibinfo
  {author} {\bibfnamefont {R.}~\bibnamefont {Zhang}}, \bibinfo {author}
  {\bibfnamefont {Z.}~\bibnamefont {Zhang}}, \bibinfo {author} {\bibfnamefont
  {H.}~\bibnamefont {Li}}, \bibinfo {author} {\bibfnamefont {Y.}~\bibnamefont
  {Wang}}, \bibinfo {author} {\bibfnamefont {G.}~\bibnamefont {Deng}}, \bibinfo
  {author} {\bibfnamefont {L.}~\bibnamefont {You}}, \bibinfo {author}
  {\bibfnamefont {H.}~\bibnamefont {Song}}, \bibinfo {author} {\bibfnamefont
  {Z.}~\bibnamefont {Wang}},  \emph {et~al.},\ }\href@noop {} {\bibfield
  {journal} {\bibinfo  {journal} {Photonics Research}\ }\textbf {\bibinfo
  {volume} {10}},\ \bibinfo {pages} {1417} (\bibinfo {year}
  {2022})}\BibitemShut {NoStop}%
\bibitem [{\citenamefont {Cheng}\ \emph {et~al.}(2019)\citenamefont {Cheng},
  \citenamefont {Zou}, \citenamefont {Guo}, \citenamefont {Wang}, \citenamefont
  {Han},\ and\ \citenamefont {Tang}}]{cheng2019broadband}%
  \BibitemOpen
  \bibfield  {author} {\bibinfo {author} {\bibfnamefont {R.}~\bibnamefont
  {Cheng}}, \bibinfo {author} {\bibfnamefont {C.-L.}\ \bibnamefont {Zou}},
  \bibinfo {author} {\bibfnamefont {X.}~\bibnamefont {Guo}}, \bibinfo {author}
  {\bibfnamefont {S.}~\bibnamefont {Wang}}, \bibinfo {author} {\bibfnamefont
  {X.}~\bibnamefont {Han}}, \ and\ \bibinfo {author} {\bibfnamefont {H.~X.}\
  \bibnamefont {Tang}},\ }\href@noop {} {\bibfield  {journal} {\bibinfo
  {journal} {Nature Communications}\ }\textbf {\bibinfo {volume} {10}},\
  \bibinfo {pages} {1} (\bibinfo {year} {2019})}\BibitemShut {NoStop}%
\bibitem [{\citenamefont {Imany}\ \emph {et~al.}(2020)\citenamefont {Imany},
  \citenamefont {Lingaraju}, \citenamefont {Alshaykh}, \citenamefont {Leaird},\
  and\ \citenamefont {Weiner}}]{imany2020probing}%
  \BibitemOpen
  \bibfield  {author} {\bibinfo {author} {\bibfnamefont {P.}~\bibnamefont
  {Imany}}, \bibinfo {author} {\bibfnamefont {N.~B.}\ \bibnamefont
  {Lingaraju}}, \bibinfo {author} {\bibfnamefont {M.~S.}\ \bibnamefont
  {Alshaykh}}, \bibinfo {author} {\bibfnamefont {D.~E.}\ \bibnamefont
  {Leaird}}, \ and\ \bibinfo {author} {\bibfnamefont {A.~M.}\ \bibnamefont
  {Weiner}},\ }\href@noop {} {\bibfield  {journal} {\bibinfo  {journal}
  {Science Advances}\ }\textbf {\bibinfo {volume} {6}},\ \bibinfo {pages}
  {eaba8066} (\bibinfo {year} {2020})}\BibitemShut {NoStop}%
\bibitem [{\citenamefont {Liu}\ \emph {et~al.}(2022)\citenamefont {Liu},
  \citenamefont {Liu}, \citenamefont {Xue}, \citenamefont {Wang}, \citenamefont
  {Li}, \citenamefont {Feng}, \citenamefont {Liu}, \citenamefont {Cui},
  \citenamefont {Wang}, \citenamefont {You} \emph {et~al.}}]{liu202240}%
  \BibitemOpen
  \bibfield  {author} {\bibinfo {author} {\bibfnamefont {X.}~\bibnamefont
  {Liu}}, \bibinfo {author} {\bibfnamefont {J.}~\bibnamefont {Liu}}, \bibinfo
  {author} {\bibfnamefont {R.}~\bibnamefont {Xue}}, \bibinfo {author}
  {\bibfnamefont {H.}~\bibnamefont {Wang}}, \bibinfo {author} {\bibfnamefont
  {H.}~\bibnamefont {Li}}, \bibinfo {author} {\bibfnamefont {X.}~\bibnamefont
  {Feng}}, \bibinfo {author} {\bibfnamefont {F.}~\bibnamefont {Liu}}, \bibinfo
  {author} {\bibfnamefont {K.}~\bibnamefont {Cui}}, \bibinfo {author}
  {\bibfnamefont {Z.}~\bibnamefont {Wang}}, \bibinfo {author} {\bibfnamefont
  {L.}~\bibnamefont {You}},  \emph {et~al.},\ }\href@noop {} {\bibfield
  {journal} {\bibinfo  {journal} {PhotoniX}\ }\textbf {\bibinfo {volume} {3}},\
  \bibinfo {pages} {1} (\bibinfo {year} {2022})}\BibitemShut {NoStop}%
\bibitem [{\citenamefont {Wen}\ \emph {et~al.}(2022)\citenamefont {Wen},
  \citenamefont {Chen}, \citenamefont {Lu}, \citenamefont {Yan}, \citenamefont
  {Xue}, \citenamefont {Zhang}, \citenamefont {Lu}, \citenamefont {Zhu},\ and\
  \citenamefont {Ma}}]{wen2022realizing}%
  \BibitemOpen
  \bibfield  {author} {\bibinfo {author} {\bibfnamefont {W.}~\bibnamefont
  {Wen}}, \bibinfo {author} {\bibfnamefont {Z.}~\bibnamefont {Chen}}, \bibinfo
  {author} {\bibfnamefont {L.}~\bibnamefont {Lu}}, \bibinfo {author}
  {\bibfnamefont {W.}~\bibnamefont {Yan}}, \bibinfo {author} {\bibfnamefont
  {W.}~\bibnamefont {Xue}}, \bibinfo {author} {\bibfnamefont {P.}~\bibnamefont
  {Zhang}}, \bibinfo {author} {\bibfnamefont {Y.}~\bibnamefont {Lu}}, \bibinfo
  {author} {\bibfnamefont {S.}~\bibnamefont {Zhu}}, \ and\ \bibinfo {author}
  {\bibfnamefont {X.-s.}\ \bibnamefont {Ma}},\ }\href@noop {} {\bibfield
  {journal} {\bibinfo  {journal} {Physical Review Applied}\ }\textbf {\bibinfo
  {volume} {18}},\ \bibinfo {pages} {024059} (\bibinfo {year}
  {2022})}\BibitemShut {NoStop}%
\bibitem [{\citenamefont {Wengerowsky}\ \emph {et~al.}(2018)\citenamefont
  {Wengerowsky}, \citenamefont {Joshi}, \citenamefont {Steinlechner},
  \citenamefont {H{\"u}bel},\ and\ \citenamefont
  {Ursin}}]{wengerowsky2018entanglement}%
  \BibitemOpen
  \bibfield  {author} {\bibinfo {author} {\bibfnamefont {S.}~\bibnamefont
  {Wengerowsky}}, \bibinfo {author} {\bibfnamefont {S.~K.}\ \bibnamefont
  {Joshi}}, \bibinfo {author} {\bibfnamefont {F.}~\bibnamefont {Steinlechner}},
  \bibinfo {author} {\bibfnamefont {H.}~\bibnamefont {H{\"u}bel}}, \ and\
  \bibinfo {author} {\bibfnamefont {R.}~\bibnamefont {Ursin}},\ }\href@noop {}
  {\bibfield  {journal} {\bibinfo  {journal} {Nature}\ }\textbf {\bibinfo
  {volume} {564}},\ \bibinfo {pages} {225} (\bibinfo {year}
  {2018})}\BibitemShut {NoStop}%
\bibitem [{\citenamefont {Joshi}\ \emph {et~al.}(2020)\citenamefont {Joshi},
  \citenamefont {Aktas}, \citenamefont {Wengerowsky}, \citenamefont
  {Lon{\v{c}}ari{\'c}}, \citenamefont {Neumann}, \citenamefont {Liu},
  \citenamefont {Scheidl}, \citenamefont {Lorenzo}, \citenamefont {Samec},
  \citenamefont {Kling} \emph {et~al.}}]{joshi2020trusted}%
  \BibitemOpen
  \bibfield  {author} {\bibinfo {author} {\bibfnamefont {S.~K.}\ \bibnamefont
  {Joshi}}, \bibinfo {author} {\bibfnamefont {D.}~\bibnamefont {Aktas}},
  \bibinfo {author} {\bibfnamefont {S.}~\bibnamefont {Wengerowsky}}, \bibinfo
  {author} {\bibfnamefont {M.}~\bibnamefont {Lon{\v{c}}ari{\'c}}}, \bibinfo
  {author} {\bibfnamefont {S.~P.}\ \bibnamefont {Neumann}}, \bibinfo {author}
  {\bibfnamefont {B.}~\bibnamefont {Liu}}, \bibinfo {author} {\bibfnamefont
  {T.}~\bibnamefont {Scheidl}}, \bibinfo {author} {\bibfnamefont {G.~C.}\
  \bibnamefont {Lorenzo}}, \bibinfo {author} {\bibfnamefont
  {{\v{Z}}.}~\bibnamefont {Samec}}, \bibinfo {author} {\bibfnamefont
  {L.}~\bibnamefont {Kling}},  \emph {et~al.},\ }\href@noop {} {\bibfield
  {journal} {\bibinfo  {journal} {Science Advances}\ }\textbf {\bibinfo
  {volume} {6}},\ \bibinfo {pages} {eaba0959} (\bibinfo {year}
  {2020})}\BibitemShut {NoStop}%
\bibitem [{\citenamefont {Guo}\ \emph {et~al.}(2018)\citenamefont {Guo},
  \citenamefont {Shi}, \citenamefont {Wang}, \citenamefont {Yang},
  \citenamefont {Ding}, \citenamefont {Ou},\ and\ \citenamefont
  {Zhao}}]{guo2018generation}%
  \BibitemOpen
  \bibfield  {author} {\bibinfo {author} {\bibfnamefont {K.}~\bibnamefont
  {Guo}}, \bibinfo {author} {\bibfnamefont {X.}~\bibnamefont {Shi}}, \bibinfo
  {author} {\bibfnamefont {X.}~\bibnamefont {Wang}}, \bibinfo {author}
  {\bibfnamefont {J.}~\bibnamefont {Yang}}, \bibinfo {author} {\bibfnamefont
  {Y.}~\bibnamefont {Ding}}, \bibinfo {author} {\bibfnamefont {H.}~\bibnamefont
  {Ou}}, \ and\ \bibinfo {author} {\bibfnamefont {Y.}~\bibnamefont {Zhao}},\
  }\href@noop {} {\bibfield  {journal} {\bibinfo  {journal} {Photonics
  Research}\ }\textbf {\bibinfo {volume} {6}},\ \bibinfo {pages} {587}
  (\bibinfo {year} {2018})}\BibitemShut {NoStop}%
\bibitem [{\citenamefont {Pavesi}\ \emph {et~al.}(2016)\citenamefont {Pavesi},
  \citenamefont {Lockwood} \emph {et~al.}}]{pavesi2016silicon}%
  \BibitemOpen
  \bibfield  {author} {\bibinfo {author} {\bibfnamefont {L.}~\bibnamefont
  {Pavesi}}, \bibinfo {author} {\bibfnamefont {D.~J.}\ \bibnamefont
  {Lockwood}},  \emph {et~al.},\ }in\ \href@noop {} {\emph {\bibinfo
  {booktitle} {Topics in applied physics}}},\ Vol.\ \bibinfo {volume} {122}\
  (\bibinfo  {publisher} {Springer},\ \bibinfo {year} {2016})\ pp.\ \bibinfo
  {pages} {1--36}\BibitemShut {NoStop}%
\bibitem [{\citenamefont {Brasch}\ \emph {et~al.}(2016)\citenamefont {Brasch},
  \citenamefont {Geiselmann}, \citenamefont {Herr}, \citenamefont {Lihachev},
  \citenamefont {Pfeiffer}, \citenamefont {Gorodetsky},\ and\ \citenamefont
  {Kippenberg}}]{brasch2016photonic}%
  \BibitemOpen
  \bibfield  {author} {\bibinfo {author} {\bibfnamefont {V.}~\bibnamefont
  {Brasch}}, \bibinfo {author} {\bibfnamefont {M.}~\bibnamefont {Geiselmann}},
  \bibinfo {author} {\bibfnamefont {T.}~\bibnamefont {Herr}}, \bibinfo {author}
  {\bibfnamefont {G.}~\bibnamefont {Lihachev}}, \bibinfo {author}
  {\bibfnamefont {M.~H.}\ \bibnamefont {Pfeiffer}}, \bibinfo {author}
  {\bibfnamefont {M.~L.}\ \bibnamefont {Gorodetsky}}, \ and\ \bibinfo {author}
  {\bibfnamefont {T.~J.}\ \bibnamefont {Kippenberg}},\ }\href@noop {}
  {\bibfield  {journal} {\bibinfo  {journal} {Science}\ }\textbf {\bibinfo
  {volume} {351}},\ \bibinfo {pages} {357} (\bibinfo {year}
  {2016})}\BibitemShut {NoStop}%
\bibitem [{\citenamefont {Engin}\ \emph {et~al.}(2013)\citenamefont {Engin},
  \citenamefont {Bonneau}, \citenamefont {Natarajan}, \citenamefont {Clark},
  \citenamefont {Tanner}, \citenamefont {Hadfield}, \citenamefont {Dorenbos},
  \citenamefont {Zwiller}, \citenamefont {Ohira}, \citenamefont {Suzuki} \emph
  {et~al.}}]{engin2013photon}%
  \BibitemOpen
  \bibfield  {author} {\bibinfo {author} {\bibfnamefont {E.}~\bibnamefont
  {Engin}}, \bibinfo {author} {\bibfnamefont {D.}~\bibnamefont {Bonneau}},
  \bibinfo {author} {\bibfnamefont {C.~M.}\ \bibnamefont {Natarajan}}, \bibinfo
  {author} {\bibfnamefont {A.~S.}\ \bibnamefont {Clark}}, \bibinfo {author}
  {\bibfnamefont {M.~G.}\ \bibnamefont {Tanner}}, \bibinfo {author}
  {\bibfnamefont {R.~H.}\ \bibnamefont {Hadfield}}, \bibinfo {author}
  {\bibfnamefont {S.~N.}\ \bibnamefont {Dorenbos}}, \bibinfo {author}
  {\bibfnamefont {V.}~\bibnamefont {Zwiller}}, \bibinfo {author} {\bibfnamefont
  {K.}~\bibnamefont {Ohira}}, \bibinfo {author} {\bibfnamefont
  {N.}~\bibnamefont {Suzuki}},  \emph {et~al.},\ }\href@noop {} {\bibfield
  {journal} {\bibinfo  {journal} {Optics express}\ }\textbf {\bibinfo {volume}
  {21}},\ \bibinfo {pages} {27826} (\bibinfo {year} {2013})}\BibitemShut
  {NoStop}%
\end{thebibliography}%

\clearpage
\setcounter{table}{0}   
\setcounter{figure}{0}   
\setcounter{equation}{0}   
\renewcommand{\thefigure}{S\arabic{figure}}
\renewcommand{\theequation}{S\arabic{equation}}
\renewcommand{\thetable}{S\arabic{table}}
\begin{widetext}
\begin{center}
\textbf{Supplementary Materials: High-quality multi-wavelength quantum light sources on silicon nitride micro-ring chip}
\end{center}
\end{widetext}

\begin{center}
\textbf{Note1: Generation and emission of correlated photon pairs with micro-ring resonator}
\end{center}
In this section, we take an all-pass-type micro-ring resonator (MRR) as an example to analyze the generation and emission of correlated photons via spontaneous four wave-mixing (SFWM) process in a ring cavity.

Figure~{\ref{fig:FigS1}} shows the schematic of an MRR consisting of a bus waveguide and a ring cavity.~The pump light at a resonant wavelength is launched into the ring cavity through the bus waveguide with a ratio of $\kappa^2$, where $\kappa^2$ is the coupling coefficient between the bus waveguide and the ring cavity. Residual pump light is straightly transmitted through the bus waveguide with a ratio of $t^2$, where $t^2$ is the transmission coefficient and satisfies $t^2+\kappa^2=1$. Via the SFWM process, signal and idler photon pairs are generated by annihilating two pump photons.
\begin{figure}[H]
    \centering
    \includegraphics[width=8.5 cm]{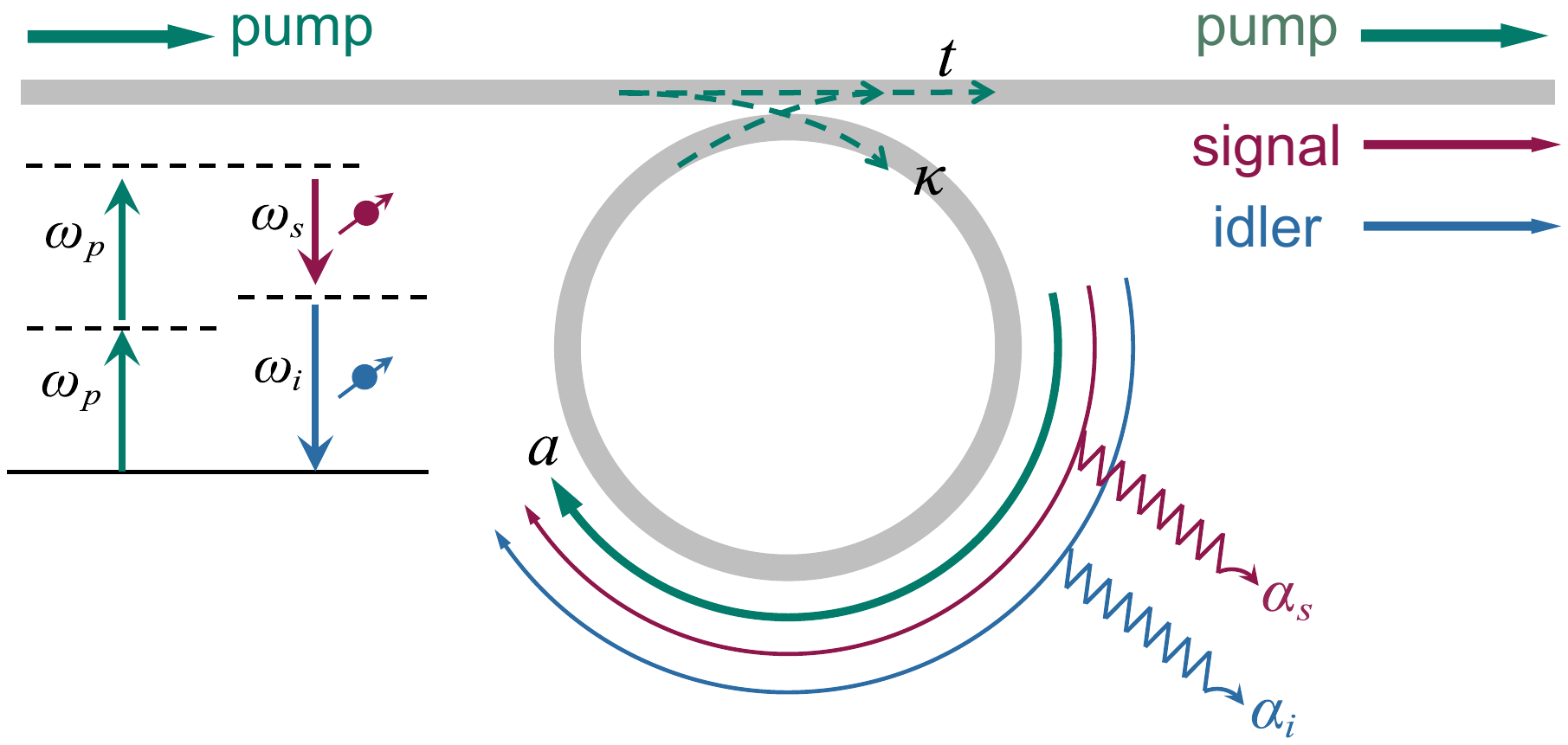}
    \caption{Schematic of an MRR consisting of a bus waveguide and a ring cavity. Inset: spontaneous four-wave mixing process in the third-order nonlinear optical material.}
    \label{fig:FigS1}
\end{figure}

The generation rate of photon pairs depends on the Q-factor of the ring cavity.~The total (loaded) Q-factor can be expressed as Eq.~{(\ref{eq:S1})}, in which $Q_i$ is the intrinsic quality factor and $Q_e$ is the external quality factor.
\begin{equation}
\label{eq:S1}
\frac{1}{Q}=\frac{1}{Q_i}+\frac{1}{Q_e}
\end{equation}

In Eq.~{(\ref{eq:S1})}, $Q_i$ and $Q_e$ can be given by the roundtrip loss ($\alpha$) in the ring cavity and the coupling efficiency ($\kappa ^2$) between the waveguide and the ring cavity,
\begin{equation}
\label{eq:S2}
Q_i=\frac{\omega _0}{\alpha \upsilon _g},
\end{equation}
\begin{equation}
\label{eq:S3}
Q_e=\frac{\omega _0}{\kappa ^2}\cdot \frac{L}{v_g},
\end{equation}
where $\omega _0$, $v_g$, and $L$ denote the angular frequency on-resonance, the light group velocity, and the roundtrip length of the cavity, respectively.

The generation rate of correlated photons via SFWM processing is given by Eq.~{(\ref{eq:S4})} \cite{guo2018generation},
\begin{equation}
\label{eq:S4}
N_c=\frac{32v_{g}^{4}\gamma ^2P_{p}^{2}Q^7}{\omega _{0}^{3}L^2Q_{e}^{4}}\cdot sinc^{2}(\frac{L\Delta \kappa}{2}),
\end{equation}
assuming that the Q-factors are wavelength-independent, where $\gamma$, $P_p$, and $\Delta\kappa$ denote the nonlinear coefficient, pump power, and the phase mismatch, respectively.~From Eq.~{(\ref{eq:S4})}, it can be inferred that the highest photon pairs generation rate can be obtained with an over-coupling condition of $Q_e/Q_i=3/4$.
\begin{figure*}
    \centering
    \includegraphics[width=18 cm]{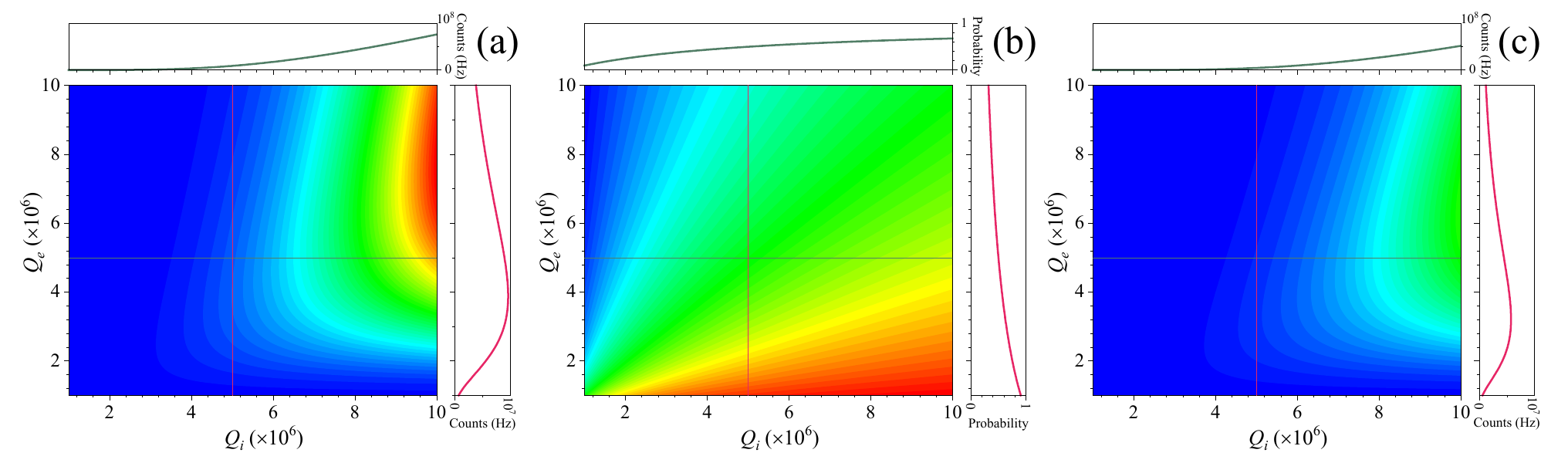}
    \caption{(a) Generation rate, (b) emission probability, and (c) emission rate of photon pairs versus $Q_e$ and $Q_i$. The green and red contour lines denotes the results for fixed $Q_e$ and $Q_i$ of $5\times10^6$, respectively.}
    \label{fig:FigS2}
\end{figure*}

However, when we consider that how many generated photon pairs can be emitted from the cavity through the bus waveguide,~such a structure with $Q_e/Q_i=3/4$ is not optimal.~Given photon pairs are generated in the ring cavity, each photon has a output probability of $\kappa^2$ per round trip, and has a probability of $\kappa^2t^2$ to enter next round trip, and then has a probability of $t^2a^2\kappa ^2$ to output, where $a^2$ represents the loss of round trip,~i.e. $a^2=e^{-\alpha \cdot L}$\cite{pavesi2016silicon}.~The probability for each photon emitting from the device is given by Eq.~{(\ref{eq:S5})}.
\begin{equation}
\begin{aligned}
\label{eq:S5}
p=&\kappa ^2+\left( 1-\kappa ^2 \right) \cdot a^2\kappa ^2\\
+&\left( 1-\kappa ^2 \right) \cdot \left( 1-\kappa ^2 \right) a^2\cdot a^2\kappa ^2+...\\
=&\frac{\kappa ^2}{1-a^2+a^2\kappa ^2}
\end{aligned}
\end{equation}

Assuming that the micro-ring has a high Q-factor with very low roundtrip loss ($a\approx1$), we can simplify $1-a^2=1-e^{\omega_0 L/v_g Q_i}$ as $\omega_0 L/v_g Q_i$, and $a^2\kappa ^2$ can be simplified as $\kappa ^2$. Thus the probability can be written as
\begin{equation}
\label{eq:S6}
p=\frac{Q}{Q_e}.
\end{equation}

The emission rate of photons $N_{cc}$ from the cavity could be expressed with phase-matching condition,~i.e. $\sin c^2\left( \frac{L\varDelta k}{2} \right) =1$.

\begin{equation}
\label{eq:S7}
N_{cc}=N_c\cdot p=\frac{32v_{g}^{4}\gamma ^2P_{p}^{2}Q^8}{\omega _{0}^{3}L^2Q_{e}^{3}}.
\end{equation}

According to Eq.~{(\ref{eq:S4})}, Eq.~{(\ref{eq:S6})}, and Eq.~{(\ref{eq:S7})}, we calculate $N_{c}$, $p$, and $N_{cc}$ with both $Q_i$ and $Q_e$ for a silicon nitride MRR with a radius of 113 $\mu$m, a width-height cross-section of 1.8 $\mu$m$\times$0.8 $\mu$m.~Via finite element method, the group velocity~$v_g$ at a pump wavelength of 1540.5 nm is $1.42\times 10^8\,\,\mathrm{m}/\mathrm{s}$, and the nonlinear coefficient $\gamma$ is $0.88~\mathrm{m}^{-1}\mathrm{W}^{-1}$ (See more details in Supplementary Materials Note2).~The incident pump power $P_p$ is 1 mW, and the pump wavelength is on resonance. As shown in Fig.~{\ref{fig:FigS2}}(a), for a given $Q_e$, the generation rate of photons increases with $Q_i$.~However, for a fixed $Q_i$, the initial increase of $Q_e$ makes $N_c$ increasing until the increase of $Q_{e}^{4}$ in Eq.~{(\ref{eq:S7})} has a greater affection on $Q^7$, as similar results given in Ref.~\cite{guo2018generation}.~It can be seen that the highest $N_c$, the optimized $Q_e$ for a given $Q_i$ appears with a slope of 3/4. Figure~{\ref{fig:FigS2}}(b) gives the emission probability from the ring cavity to the bus waveguide with different $Q_i$ and $Q_e$, which illustrates that the probability increases with $Q_i$ while decreases with $Q_e$.~Therefore, as shown in Fig.~{\ref{fig:FigS2}}(c), the highest $N_{cc}$ reaches with a slope of 3/5 for the optimized ratio between $Q_e$ and $Q_i$. From our calculation, the generation rate of photon pairs increases with the $Q_i$ of the cavity under the same pump power.~The cavity with high $Q_i$ can be achieved by reducing the scattering loss, bending loss, and nonlinear loss during the fabrication processes.~And considering the emission processes of the photon pairs, an optimized $Q_e$ should be selected by designing the gap width between the bus waveguide and the ring cavity.

\maketitle
\begin{center}
\textbf{Note2: Device fabrication and characterization}
\end{center}
Our micro-ring resonators are fabricated by LIGETEC using AN800 technology with high optical quality silicon oxides and high optical quality stoichiometric silicon nitride.~A waveguide with a thickness of 800 nm is fully cladded and embedded with optical quality oxide.~For correlated photons generation, one requirement is the conservation of momentum, which needs the group velocity dispersion (GVD) of the devices to be zero and appropriately anomalous.~In our experiments, we choose the width-height cross-section of the waveguide with 1.8 $\mu$m$\times$0.8 $\mu$m based on our theoretical calculation, as shown in the inset of Fig.~{\ref{fig:FigS3}}(a).~For such a micro-ring resonator, the calculated effective refractive index $n_{eff}$ of $TE_{00}$ and $TM_{00}$ mode are illustrated in Fig.~{\ref{fig:FigS3}}(a) and the GVD are shown in Fig.~{\ref{fig:FigS3}}(b).~It can be seen that near-zero anomalous-GVD with a large bandwidth is reached around 1550 nm.~At a pump light at 1540.5 nm in our experiments, the effective refractive index of $TE_{00}$ and $TM_{00}$ are 1.85 and 1.82, and the GVDs are $-8.48\times10^{-26}$ and $-4.5\times10^{-26}~\mathrm{s}^2/\mathrm{m}$, respectively.~The obtained mode profiles of $TE_{00}$ and $TM_{00}$ are shown in the inset of Fig.~{\ref{fig:FigS3}}(b).~The effective mode field areas $A_{eff}$ are 1.16 and 1.08 $\mu m^2$ for $TE_{00}$ and $TM_{00}$, respectively.~The third-order nonlinear optical coefficient $\gamma$ of such a silicon nitride waveguide is 0.88 and 0.82~$\mathrm{W}^{-1}\mathrm{m}^{-1}$ for $TE_{00}$ and $TM_{00}$, respectively, where $\gamma$ is defined as $\gamma =2\pi n_2/\left(\lambda A_{eff} \right)$ and the nonlinear index $n_2$ is $0.25\times10^{-18}~\mathrm{m}^2/\mathrm{W}$ for silicon nitride.

\begin{figure}
    \centering
    \includegraphics[width=8.5 cm]{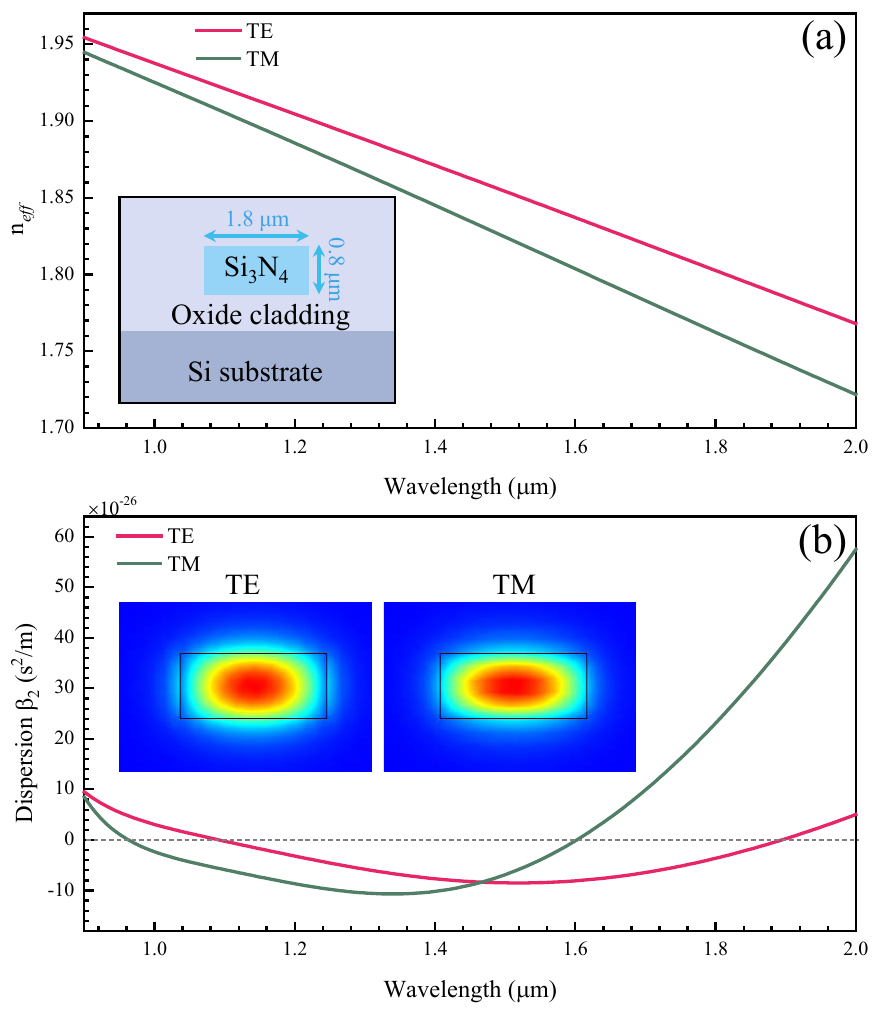}
    \caption{(a) Calculated results of the effective refractive index and (b) group velocity dispersion for $TE_{00}$ and $TM_{00}$ modes of silicon nitride waveguide with a width-height cross-section of 1.8 $\mu$m$\times$0.8 $\mu$m as inserted in Fig.~{\ref{fig:FigS3}}(a). The inset in Fig.~{\ref{fig:FigS3}}(b) shows the mode profiles of $TE_{00}$ and $TM_{00}$.}
    \label{fig:FigS3}
\end{figure}
To be compatible with the ITU wavelength of fiber communication,
~we use a device with a free spectral range (FSR) of 200 GHz. According to Eq.~{(\ref{eq:S8})}, its radius is 113~$\mu$m.
\begin{equation}
\label{eq:S8}
R=\frac{c}{n_g\cdot 2\pi \cdot \mathrm{FSR}},
\end{equation}
where $c$ is the speed of light in vacuum, and $n_g$ is the group refractive index which is defined as $n_{\mathrm{g}}=n_{eff}-\lambda _0dn_{eff}/d\lambda$, respectively.

To optimize the generation and emission of correlated photons, the gap width between the bus waveguide and the ring cavity is selected.~We first calculate the transmission of a micro-ring with a gap width of 0.15 $\mu$m.~Figure~{\ref{fig:FigS4}} shows the transmissions from 1500 nm to 1600 nm with a ring radius of 3.5 $\mu$m, which corresponds to the over-coupling regime.~The inset on the left gives the distribution of the electric field for the off-resonance case, and the right one gives the on-resonance case, in which the input light passes through the bus waveguide directly and enters into the ring cavity, respectively.~By scanning the gap width, we obtain that the critical coupling requires a gap width of~0.40 $\mu$m.~It is worth noting that in the calculation, the increment of grid refinement increases the computational resource exponentially, thus we calculate a device with small radius.

Four groups of silicon nitride MRR devices with gap widths of 0.35, 0.40, 0.45, 0.50 $\mu$m are designed and fabricated by foundry, which cover the over-coupling, critical coupling, and under-coupling regimes, respectively.~As shown in Fig.~{\ref{fig:FigS5}}(a), the radius of the ring cavity is 113 $\mu$m, and the width-height cross-section of the waveguide is 1.8~$\mu$m$\times$0.8~$\mu$m. The input and output bus waveguides are featured with mode converters and are end-coupled with anti-reflection coated lensed fiber, with a coupling loss of 1.0 dB per facet.~The performance of the micro-ring resonators is measured using a tunable laser from 1480 nm to 1620 nm with a linewidth of \textless10 kHz (Toptica DLC Pro).~Figure~{\ref{fig:FigS5}}(b) gives a measured resonance dip at 1540.5 nm with a measured $Q$, $Q_e$ and $Q_i$ of 1.0$\times$10$^6$, 1.4$\times$10$^6$, and 3.5$\times$10$^6$, respectively. The transmission properties and loaded Q-factors are further measured and shown in Fig.~{\ref{fig:FigS5}}(c), indicating that the free spectral range (FSR) of the cavity is $\sim$200 GHz and the Q-factors are up to $\sim$10$^6$ for all measured resonances.~It can be seen that all the Q-factors are almost the same within the telecom band, which satisfies the assumption of Eq.~{(\ref{eq:S4})}. The resonance of another three groups of devices around 1540.5 nm are measured and listed in Table~{\ref{tab:tableSI}}. According to the fitted full width at half maximum (FWHM), the loaded quality of the four groups of devices are calculated to be 1.0$\times10^6$, 1.6$\times10^6$, 2.2$\times10^6$, and 3.1$\times10^6$ with an extinction ratio (ER) of 7.5, 12.5, 24.5, and 14.1 dB, corresponding to over-coupling, over-coupling, critical coupling, and under-coupling, respectively.

\begin{figure}
    \centering
    \includegraphics[width=8.5 cm]{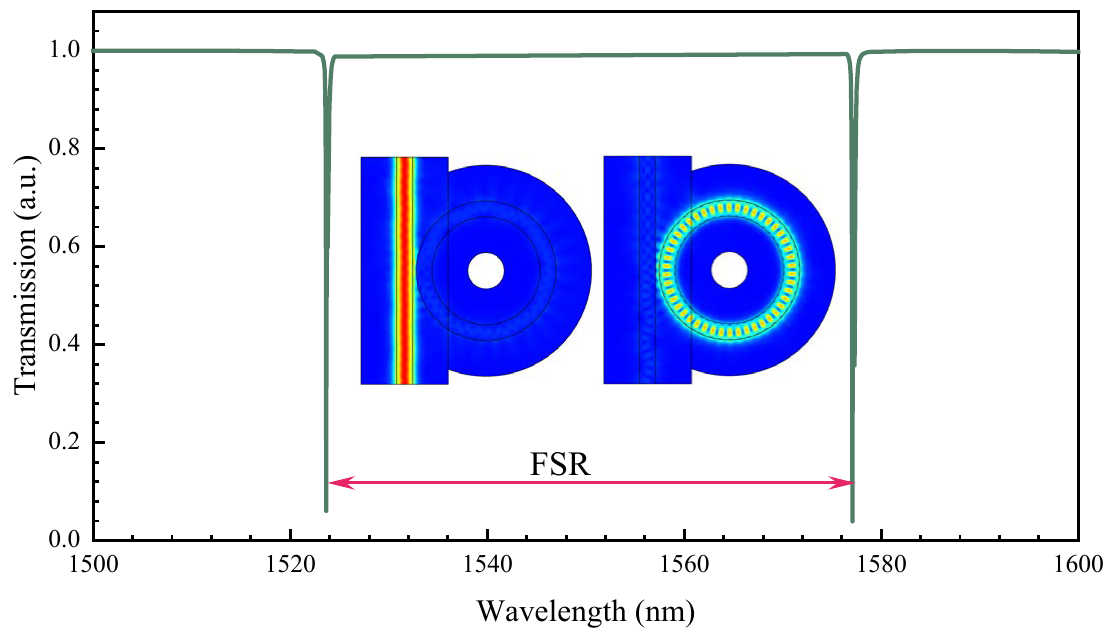}
    \caption{Simulated transmission spectrum for an over-coupling device with a gap width of 0.15 $\mu$m and a radius of 3.5 $\mu$m. Inset on the left and right show the distribution of electric field for off- and on-resonance cases, respectively.}
    \label{fig:FigS4}
\end{figure}
\begin{figure}
    \centering
    \includegraphics[width=8.5 cm]{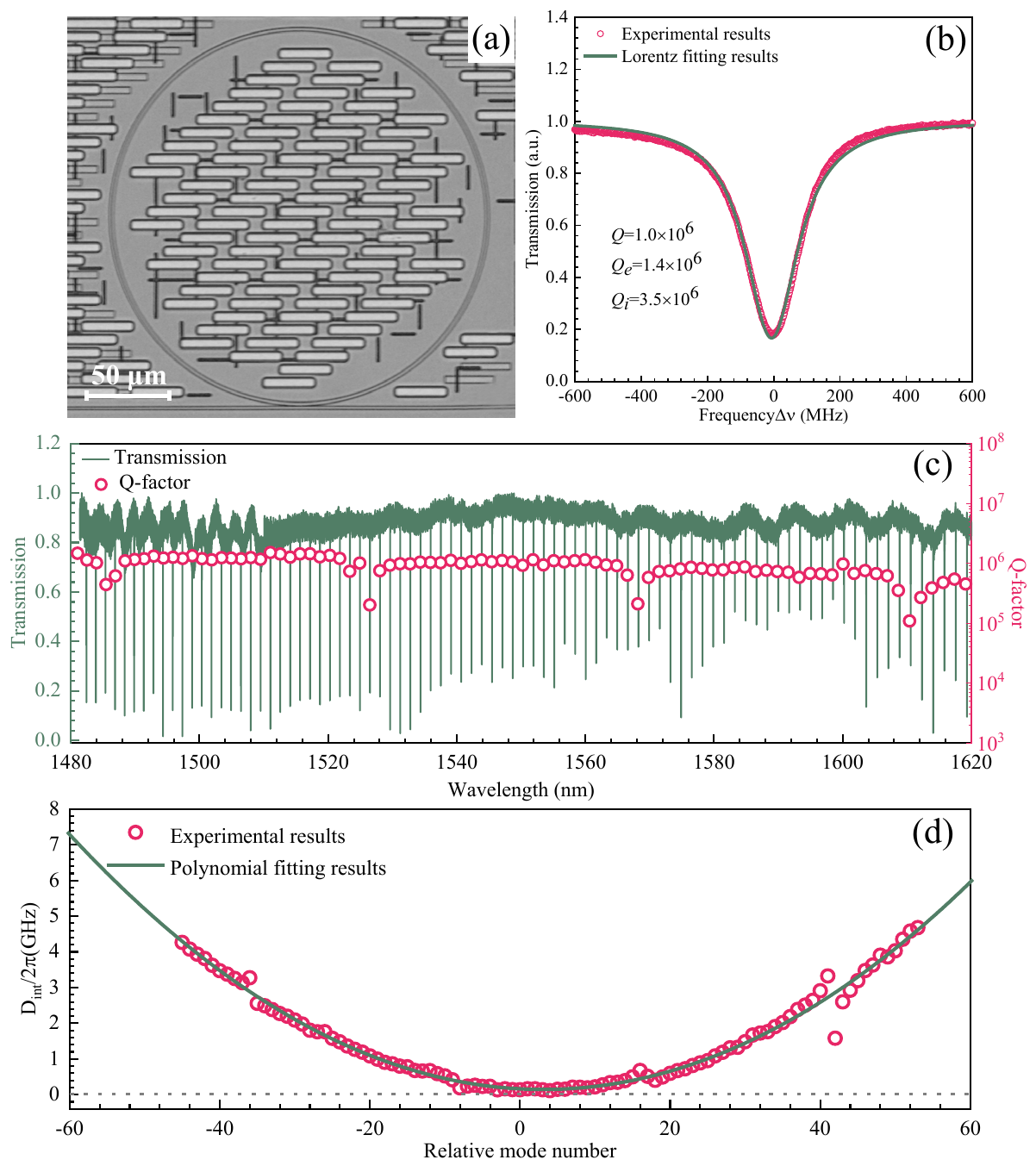}
    \caption{Device design and characterization.~(a) Microscopy image of the silicon nitride micro-ring with a thickness of 0.8 $\mu$m, width of 1.8 $\mu$m, radius of 113 $\mu$m, and gap width of 0.35 $\mu$m.~(b) Resonance dip at 1540.5 nm, with a $Q=1.0\times10^{6}$, $Q_{e}=1.4\times10^{6}$, and $Q_{i}=3.5\times10^{6}$.~(c) Transmissions and loaded Q-factors from 1480 nm to 1620 nm for 87 resonances with a free spectral range of $\sim200$ GHz and a $Q$ factor of $\sim10^6$.~(d) Integrated dispersion $D_{int}$ extracted from the device transmission and calculated based on the resonant wavelengths, suggesting the second-order GVD of $-0.88\times$10$^{-25}$~$\mathrm{s}^{2}$/$\mathrm{m}$.}
    \label{fig:FigS5}
\end{figure}

\begin{table*}
\caption{\label{tab:tableSI} Resonance details of four groups of devices.}
\begin{ruledtabular}
\begin{tabular}{ccccccc}
gap width & 3 dB linewidth & loaded Q & ER (dB) & coupling &$Q_e$&$Q_i$\\\hline
0.35 $\mu$m &193 MHz & 1.0$\times10^6$ & 7.5 & over & 1.4$\times10^6$&3.5$\times10^6$\\
0.40 $\mu$m &123 MHz & 1.6$\times10^6$ & 12.5 & over &2.6$\times10^6$ &4.1$\times10^6$\\
0.45 $\mu$m &90 MHz & 2.2$\times10^6$ & 24.5 & critical &4.1$\times10^6$ &4.6$\times10^6$\\
0.50 $\mu$m &62 MHz & 3.1$\times10^6$ & 14.1 & under & 7.8$\times10^6$&5.3$\times10^6$\\
\end{tabular}
\end{ruledtabular}
\end{table*}

We measure the dispersion property of the micro-ring resonators following the integrated dispersion approach\cite{brasch2016photonic}.~The cavity resonant frequency $\omega _{\mu}$ is extracted by using a wave meter (WSU-10, High Finesse) with a precision of 10 MHz and a resolution of 0.4 MHz.~The Taylor series is used to expand the resonant frequency with a respective mode index $\mu$ as expressed in Eq.~{(\ref{eq:S9})}, where $\mu =0$ corresponds to the pump light mode.
\begin{equation}
\begin{aligned}
\label{eq:S9}
\omega _{\mu}=&\omega _0+D_1\mu +\frac{D_2}{2}\mu ^2+\sum_{n=3}^{\infty}{\frac{D_n}{n!}}\mu ^n
\\
=&\omega _0+D_1\mu +D_{int}
\end{aligned}
\end{equation}
In Eq.~{(\ref{eq:S9})}, $D_1$ is the FSR of the angular frequency, $D_2$ is the second-order dispersion, $D_n$ represent the high-order dispersion, and $D_{int}$ is the integrated dispersion.~In our experiment, the measured integrated dispersion $D_{int}$ of the micro-resonator is plotted in Fig.~{\ref{fig:FigS5}}(d). With a polynomial fitting of the integrated dispersion of resonant frequency, we obtain $D_2=2.28\times 10^7$~Hz.~The converted second-order GVD parameter $\beta_2$ is $-0.88\times10^{-25}$ $\mathrm{s}^2/\mathrm{m}$ according to the expression of $\beta _2=-\left(n\cdot D_2\right) /\left(c\cdot D_{1}^{2}\right)$. It reveals purely anomalous GVD, thus fulfill the requirement of phase-matching for the generation of correlated photons via the third-order nonlinear process.

\maketitle
\begin{center}
\textbf{Note3: Spectra of noise and correlated photon pairs}
\end{center}
The silicon nitride MRR is pumped by a continuous wave (CW) laser, the power of which is controlled and monitored by a variable optical attenuator (VOA) and 90/10 beam splitter (BS) with a power meter (PM).~The polarization state of the pump laser is manipulated by a polarization controller (PC).~The sideband noise of the pump light is cleaned by a dense-wavelength division multiplexer (DWDM1) with an extinction ratio of \textgreater120 dB. The MRR is connected with noise-rejecting filters (DWDM2 and DWDM3) with 10 cm fiber-pigtails.~The generated photons are detected by superconducting nanowire single photon detectors (SNSPDs, P-CS-6, PHOTEC), the outputs of which are recorded by a time-to-digital converter (TDC, ID900, ID Quantique).

To evaluate the spontaneous Raman noise and the correlated photons generated in the fiber-pigtails and the silicon nitride MRR, their spectra are recorded by frequency scanned photon counting using a tunable bandpass filters (TBF, XTA-50, EXFO) with a bandwidth of 80~GHz.~The spontaneous Raman noise generated by the fiber-pigtails without the MRR is shown in Fig.~{\ref{fig:FigS6}}(a).~The residual pump light at 1540.5 nm is negligible thanks to the high-rejection filters.~The measured spectral envelope is as the same as the Raman scattering in the fiber.~Next, we put the MRR device in our experimental setup and measure the single photon counts when the pump wavelength is off-resonance.~Figure~{\ref{fig:FigS6}}(b) shows a different spectral envelop, which could attribute to the Raman scattering from the silicon nitride device.
\begin{figure}
    \centering
    \includegraphics[width=8.5 cm]{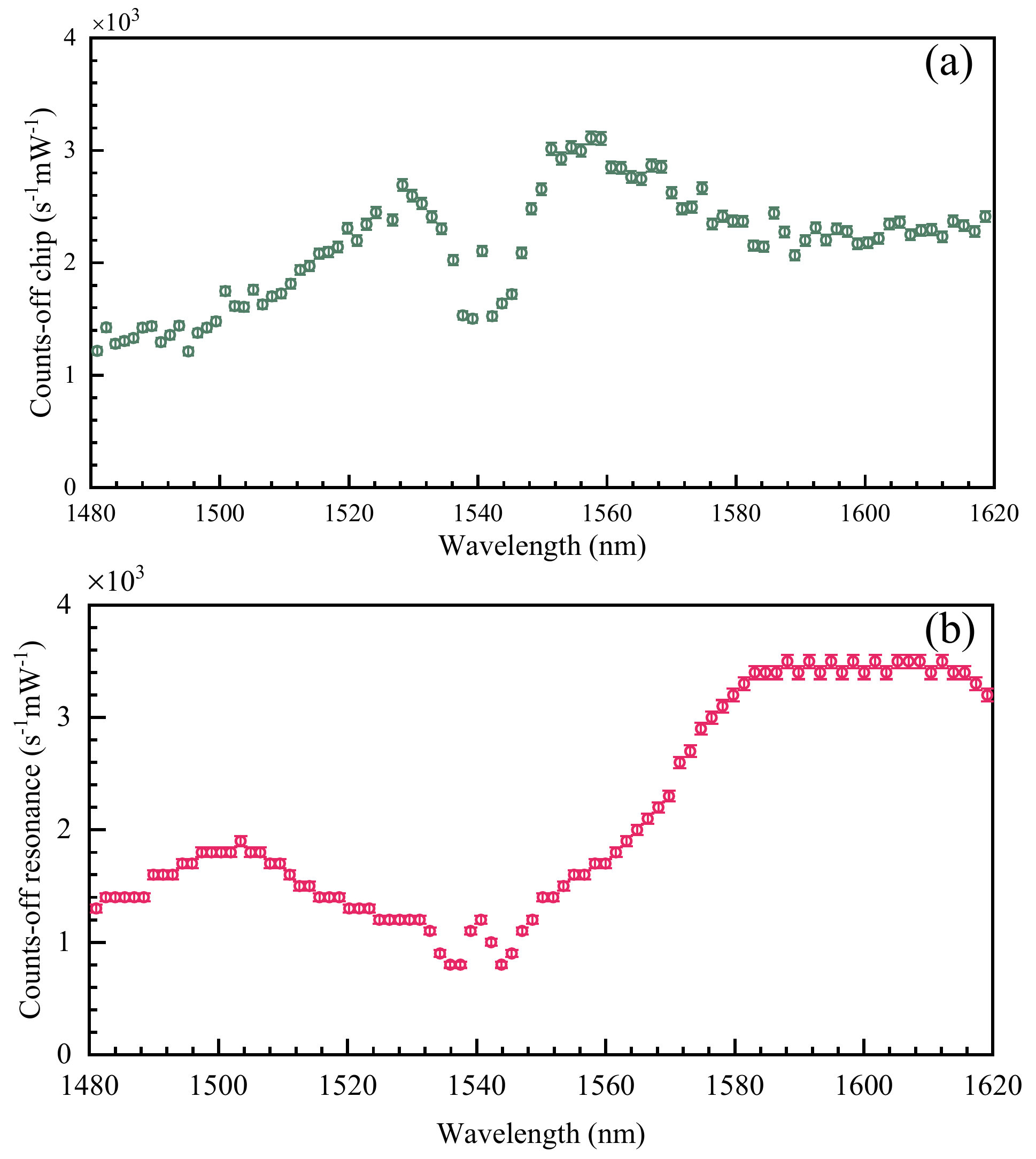}
    \caption{Spectra of noise photons. (a) Spontaneous Raman noise generated by the experimental setup without the MRR, and (b) the spectral response of the MRR for off-resonance case.}
    \label{fig:FigS6}
\end{figure}
\begin{figure}
    \centering
    \includegraphics[width=8.5 cm]{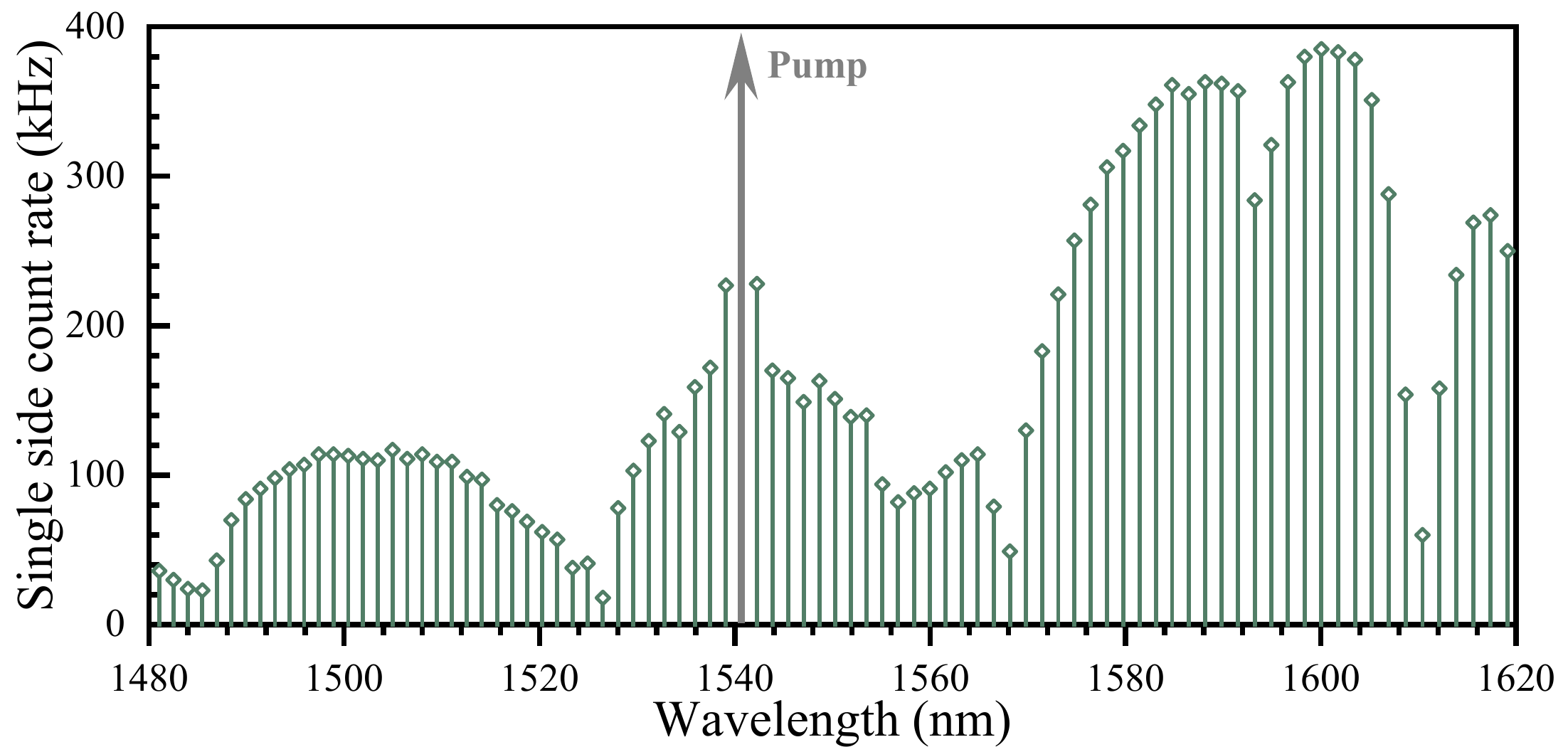}
    \caption{Spectral response of the MRR for on-resonance case with a pump power of 1 mW.}
    \label{fig:FigS7}
\end{figure}

Figure~{\ref{fig:FigS7}} illustrates the measured spectrum for the on-resonance case with a pump power of 1 mW.~We can clearly observe the comb structure associated with the resonance of MRR, which extends as far as 140 nm, i.e. with 87 frequency comb lines on both sides of pump light.~We measure each comb line for varying pump powers and extract the linear and quadratic components of the count rate. Our results show that only seven pairs of frequency comb lines have correlated photons.~Based on the analysis of Raman noise for off-chip and off-resonance cases, we can infer that most of the noise comes from silicon nitride MRR due to the Raman scattering.

\maketitle
\begin{figure}[H]
    \centering
    \includegraphics[width=8.5 cm]{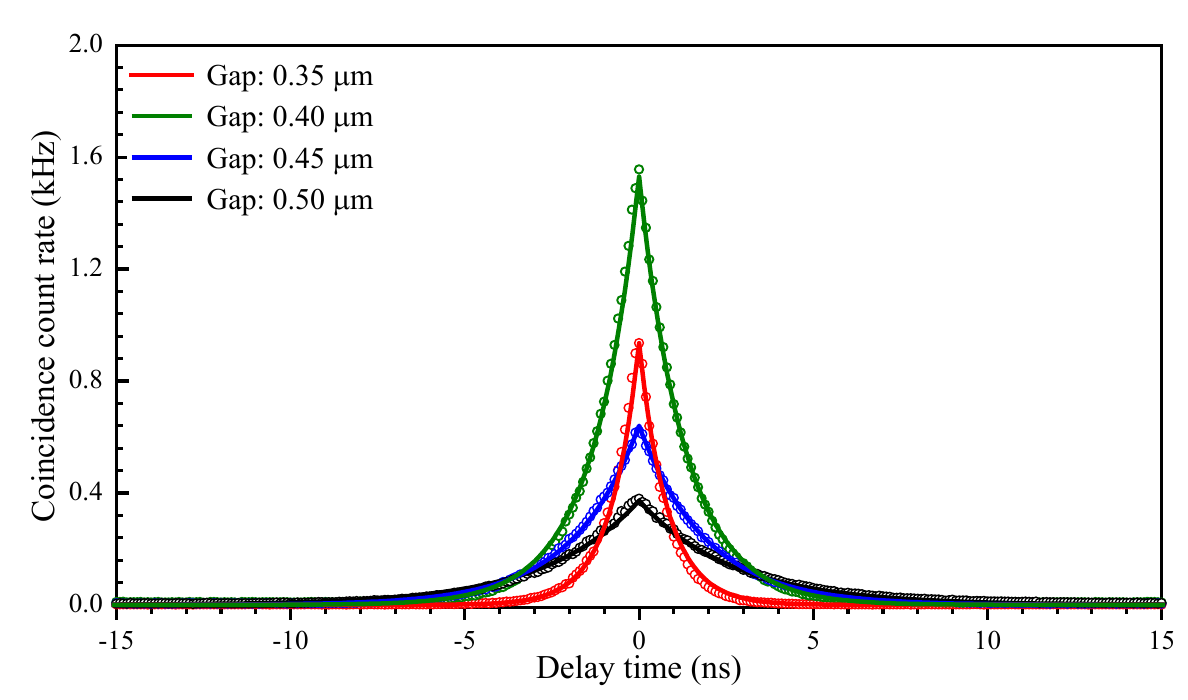}
    \caption{Coincidence histograms of four groups of devices with the gap width of 0.35, 0.40, 0.45, and 0.50 $\mu$m.}
    \label{fig:FigS8}
\end{figure}
\begin{figure}
    \centering
    \includegraphics[width=8.5 cm]{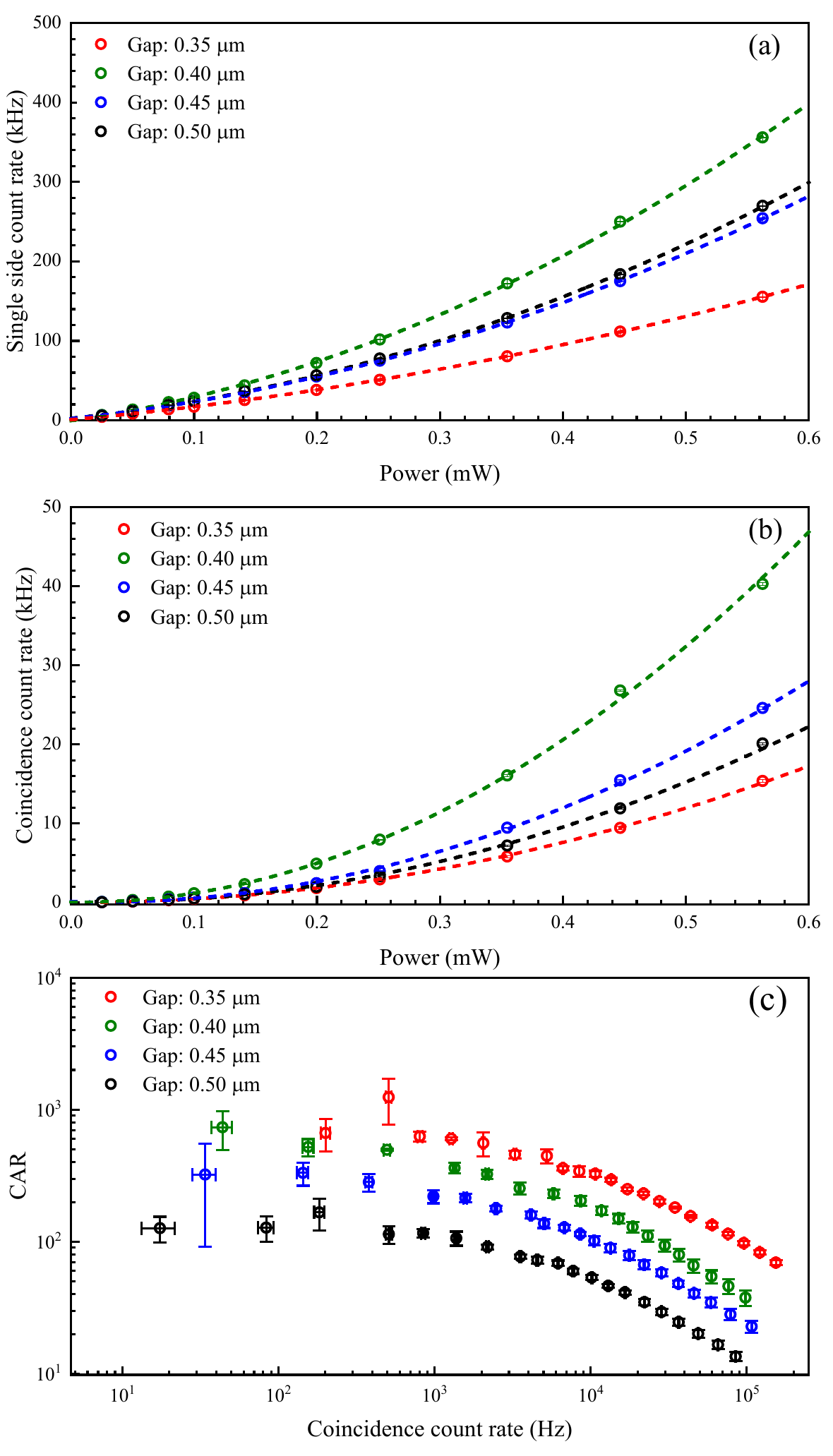}
    \caption{Quantum correlation property in different devices. (a) Single side count rate, (b) coincidence count rate, and (c) comparison of the CAR values and the detected coincidence count rate among different devices with the gap width of 0.35, 0.40, 0.45, and 0.50 $\mu$m.}
    \label{fig:FigS9}
\end{figure}

\begin{center}
\textbf{Note4: Generation and emission of photon pairs with different coupling conditions}
\end{center}
\begin{table*}
\caption{\label{tab:tableSII} Emission and transmission of four groups of devices.}
\begin{ruledtabular}
\begin{tabular}{ccccc}
  & 0.35 $\mu$m & 0.40 $\mu$m& 0.45 $\mu$m &0.50 $\mu$m\\\hline
Total collection efficiency & 22.9$\pm$0.5\% &21.5$\pm$0.7\% & 15.4$\pm$1.1\% &12.1$\pm$0.4\%\\
Coupling loss per facet/dB\footnote{The coupling loss contains the insertion loss of a DWDM that fused with the lensed fiber with 0.5 dB.} & 1.50& 1.50 & 1.45 &1.45\\
Transmission loss/dB & 2.55 &2.55 &2.55 &2.55\\
Detection loss/dB &1.10&1.10&1.10&1.10\\
Emission probability-measured & 75.0$\pm$2.1\%&69.5$\pm$3.3\%& 49.9$\pm$5.3\%&39.3$\pm$2.1\%\\
Emission probability-predicted & 71.2\%&61.9\%& 53.0\%&40.2\%\\
\end{tabular}
\end{ruledtabular}
\end{table*}
To examine the generation and emission of photon pairs, we measure the quantum correlation of another three groups of devices with different coupling conditions. At a pump power level of 0.56 mW, the histograms of the coincidence between signal and idler photons are shown in Fig.~{\ref{fig:FigS8}}.~It can be seen that the FWHM of the histogram increases with the loaded Q-factor of the MRR.~In other words, the coherence time of the photons increases with the Q-factors, in which the FWHM is 1.64, 2.64, 3.82, and 5.40 ns, respectively, and corresponds to the cavity linewidth shown in Table~{\ref{tab:tableSI}}.
~Figure~{\ref{fig:FigS9}}(a) and (b) show the measured single side count rate and coincidence count rate on the resonance of 1550.1 nm under different pump power levels, respectively.~The red, green, blue, and black circles are the experimental results for the gap width of 0.35, 0.40, 0.45, and 0.50 $\mu$m, respectively. The lines are the quadratic polynomial fitting of the experimental results.  With a coincidence time window of the FWHM of histograms calculated from Fig.~{\ref{fig:FigS8}},~Fig.~{\ref{fig:FigS9}}(c) shows the comparison of the coincidence-to-accidental ratio (CAR) and the detected photon pair rate among different groups of devices with the gap width of 0.35, 0.40, 0.45, and 0.50 $\mu$m, respectively.

Then we calculate the collection efficiencies of the photon pair with different groups of devices. According to the relationship among the single side count rate, coincidence count rate, and accidental coincidence count rate, we can obtain the following equations\cite{engin2013photon}:

\begin{equation}
\begin{aligned}
\label{eq:S10}
\left\{ \begin{array}{l}
	N_s=\eta _s(N_c+R_s)+d_s\\
	N_i=\eta _i(N_c+R_i)+d_i\\
	N_{cc}=N_c\eta _s\eta _i+N_{ac}\\
\end{array} \right.
\end{aligned}
\end{equation}
where $N_s$, $N_i$, $N_{cc}$, and $N_{ac}$ are the detected signal, idler, coincidence, and accidental coincidence count rate, $N_c$ is the photon pair generation rate, $\eta _s$ and $\eta _i$ are the collection efficiency of signal and idler photons, $R _s$ and $R _i$ are the generation rate of noise photons in the signal and idler channel, $d _s$ and $d _i$ are the dark count rate of SNSPDs in the signal and idler channels, respectively.
\begin{figure}
    \centering
    \includegraphics[width=8.5 cm]{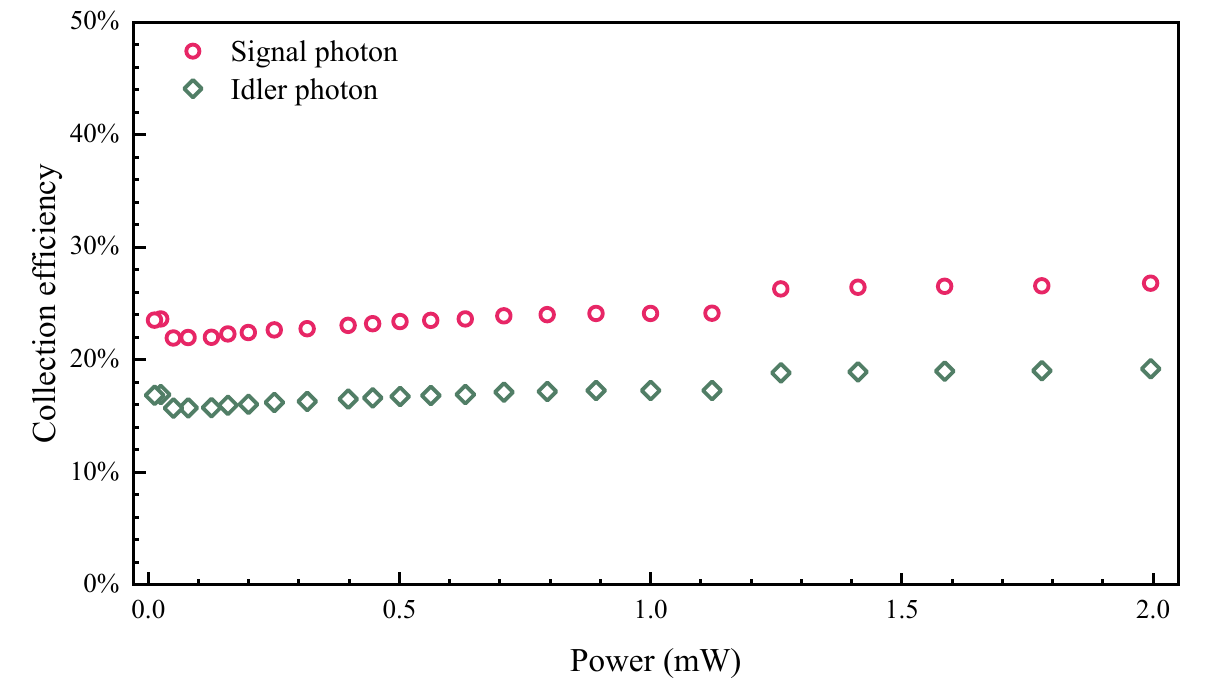}
    \caption{Collection efficiency for signal and idler photons using a MRR with a gap width of 0.35 $\mu$m.}
    \label{fig:FigS10}
\end{figure}

Figure~{\ref{fig:FigS10}} shows the collection efficiency of signal and idler photons for the device with a gap width of 0.35 $\mu$m. We obtain a collection efficiency of 22.9\% for signal photons, which contains the coupling loss, transmission losses, detection loss, and extraction loss. The first three components are directly measured in our experiment. The extraction loss, attributed to the probability $p$ emitted from the ring cavity and into the bus waveguide, can be inferred by subtracting the first three losses from the total transmission losses calculated from the collection efficiency. The calculated emission probability is $75.0\pm2.1\%$, which is consistent with the theoretical value of 71.2\% ($Q/Q_e$), which depends on the coupling conditions. We also calculate the collection efficiency of signal photons with another three groups of devices, as listed in Table~{\ref{tab:tableSII}}. It can be seen that the total collection efficiency decreases with the gap width, ranging from over-coupling to under-coupling.~Our results show that the MRRs can be designed according to Eq.~{(\ref{eq:S7})} for optimizing the performance of photon pairs source.

\end{document}